\documentclass{ap-jnmp}
\usepackage{bm}

\begin{document}

\newcommand\eref[1]{(\ref{#1})}
\renewcommand\vec[1]{\bm{#1}}
\renewcommand\le{\leqslant} 
\renewcommand\ge{\geqslant} 

\newcommand\myedge[1]{\mathsf{#1}}
\newcommand\mynode[1]{\mathsf{#1}}
\newcommand\myvar[2][0]{#2}
\newcommand\myconst[2]{#1_{#2}}
\newcommand\mytriangle{{\!\scriptscriptstyle\triangle}}
\newcommand\mymatrix[1]{\mathsf{#1}}
\newcommand\myunitket{|\,1\,\rangle}  
\newcommand\mybra[1]{\langle\,#1\,|} 
\newcommand\myShift[2][T]{\mathbb{#1}_{#2}}
\newcommand\myShifted[3][T]{ \mathop{\mathbb{#1}_{#2}}{#3}}
\newcommand\myShiftedB[3][T]{\left( \mathop{\mathbb{#1}_{#2}}{#3} \right)}
\newcommand\mytoeplitz[3]{\mathsf{#1}^{#2}_{#3}}

%%%%%%%%%%%%%%%%%%%%%%%%%%%%%%%%%%%%%%%%%%%%%%%%%%%%%%%%%%%%%%%%%%%%%%%%%%%%%%%%
%%%%%%%%%%%%% Title, author, address etc. %%%%%%%%%%%%%%%%%%%%%%%%%%%%%%%%%%%%%%
%%%%%%%%%%%%%%%%%%%%%%%%%%%%%%%%%%%%%%%%%%%%%%%%%%%%%%%%%%%%%%%%%%%%%%%%%%%%%%%%

\markboth{V.E. Vekslerchik}
{Explicit solutions for a nonlinear model on the honeycomb and triangular lattices.}

%%\catchline{}{}{2016}{}{}
%%%%%%%%%%%%%%%%%%%%%%%%%%%%%%%%%%%%%%%%%%%%%%%%%%%%%%%%%%%%%%%%%%%%

\copyrightauthor{V.E. Vekslerchik}

\title{Explicit solutions for a nonlinear model 
       on the honeycomb and triangular lattices.}

\author{V.E. Vekslerchik}

\address{
Usikov Institute for Radiophysics and Electronics \\
12, Proskura st., Kharkov, 61085, Ukraine \\
\email{vekslerchik@yahoo.com}
}

\maketitle

\begin{abstract}
We study a simple nonlinear model defined on the honeycomb and triangular 
lattices. We propose a bilinearization scheme for the field equations and 
demonstrate that the resulting system is closely related to the well-studied 
integrable models, such as the Hirota bilinear difference equation and the 
Ablowitz-Ladik system. This result is used to derive the two sets of explicit 
solutions: the $N$-soliton solutions and ones constructed of the Toeplitz 
determinants. 
\end{abstract}

\keywords{
  integrable lattice models, 
  honeycomb lattice, 
  triangular lattice,
  bilinear approach,
  explicit solutions
}

\ccode{2010 Mathematics Subject Classification: 
  37J35, %Completely integrable systems, topological structure of phase space, integration methods 
  11C20, %Matrices, determinants [See also 15B36] 
  35C08, %Soliton solutions 
  15B05  %Toeplitz, Cauchy, and related matrices 
}

%%%%%%%%%%%%%%%%%%%%%%%%%%%%%%%%%%%%%%%%%%%%%%%%%%%%%%%%%%%%%%%%%%%%%%%%%%%%%%%
\section{Introduction.}
%%%%%%%%%%%%%%%%%%%%%%%%%%%%%%%%%%%%%%%%%%%%%%%%%%%%%%%%%%%%%%%%%%%%%%%%%%%%%%%

We study a simple nonlinear model defined on the honeycomb and triangular 
lattices (HL and TL). 
The main aim of this work is to extend the direct methods of the soliton 
theory to the case of `non-square', i.e. different from $\mathbb{Z}^{2}$, 
lattices.

Although there has been considerable interest in the integrable nonlinear 
models on such lattices (see e.g. 
\cite{AB03,BMS05,DNS07,A98,A01,BS02,BHS02,BH03,AS04,BS10}) there are still 
many problems to solve in this field. 
The main source of difficulties arising in studies of 
integrable models on the HL and TL 
%honeycomb lattice (HL) 
is a lack of natural ways to 
separate variables. This hinders the usage of the standard approaches like, 
for example, the inverse scattering transform.

There are several strategies to address the non-square lattices. 
One of them, which was used, in, for example, \cite{AB03,BMS05,DNS07}, is to 
consider the HL and TL as sublattices of sections of 
the $\mathbb{Z}^{3}$-lattice. In other words, this approach is to consider the 
model in question as a restriction of a more general (higher dimensional) one. 
Another approach has been developed in the works of Adler, Bobenko, Suris and 
co-authors \cite{A98,A01,BS02,BHS02,BH03,AS04,BS10} who elaborated a framework 
for integrable models on non-square lattices and, more generally, on arbitrary 
graphs. Among different aspects of this approach we would highlight two 
moments. First, there is an almost algorithmical way to convert an integrable 
model on a graph that possesses the property of the three-dimensional 
consistency \cite{ABS03} to the one on the quad-graph \cite{A01,BS02}, 
i.e. to the system of four-point equations 
(compare with the system of the discrete Moutard equations from \cite{DNS07}). 
The second ingredient is the special form of the Lax (or zero-curvature) 
representation, which is called in \cite{A01} the ``trivial monodromy 
representation''. 
The results of \cite{A98,A01,BS02,BHS02,BH03,AS04,BS10} provide answers to many 
questions arising in the theory of integrable systems. However, if we 
consider the problem of finding solutions, we have to admit that there is 
still much work to be done. 
While in the classical integrable models (on the square lattice, in our case) 
the zero-curvature representation is a base for the inverse scattering 
transform which is a tool to derive solutions 
(at least to linearize the problem), the corresponding methods for models on 
graphs, that use the trivial monodromy representation, are, so far, to be 
developed. 
The possibility to convert an integrable model into the one on the quad-graph 
is very attracting from the viewpoint of the so-called direct methods. 
However, a practical implementation of this idea may face some difficulties 
(we return to this question in section \ref{sec-crs}).

In this work we do not study general aspects of the integrability  and the 
geometry of the HL and TL. We restrict ourselves to the following problem: to 
find explicit solutions for one of the `universal' integrable models of the 
paper \cite{A01} which was studied in \cite{BS02}. 
This model is described in section \ref{sec-def}. 
In section \ref{sec-bilin}, we introduce new variables (the tau-functions), 
bilinearize the field equations and demonstrate that the resulting system is 
closely related to the well-studied integrable models, such as the 
Hirota bilinear difference (or discrete KP) and the Ablowitz-Ladik equations 
(the important difference between our approach and the one of 
\cite{DNS07,BS02} is that we use, instead of the four-point quad-equations, 
a system of three-point equations).
Then, we present the two sets of explicit solutions for this system that 
provide the two families of explicit solutions for the equations we want to 
solve: solutions constructed of the Toeplitz determinants and the $N$-soliton 
solutions (see section \ref{sec-exact-h} for the case of HL and 
section \ref{sec-exact-t} for the case of TL).

Of course, during the implementation of this standard program, we meet the 
manifestations of the peculiar features of the HL and TL. However, as a reader 
will see, one can overcome the emerging complications by elementary means.

%%%%%%%%%%%%%%%%%%%%%%%%%%%%%%%%%%%%%%%%%%%%%%%%%%%%%%%%%%%%%%%%%%%%%%%%%%%%%%%
\section{Definitions and main equations. \label{sec-def} }
%%%%%%%%%%%%%%%%%%%%%%%%%%%%%%%%%%%%%%%%%%%%%%%%%%%%%%%%%%%%%%%%%%%%%%%%%%%%%%%

%%%%%%%%%%%%%%%%%%%%%%%%%%%%%%%%%%%%%%%%%%%%%%%%%%%%%%%%%%%%%%%%%%%%%%%%%%%%%%%
\subsection{Honeycomb lattice.}
%%%%%%%%%%%%%%%%%%%%%%%%%%%%%%%%%%%%%%%%%%%%%%%%%%%%%%%%%%%%%%%%%%%%%%%%%%%%%%%

The model that we study in this paper can be defined in terms of the 
action  
\begin{equation}
  \mathcal{S}  
  = 
  \sum\limits_{\myedge{e} \in \mathsf{E}} 
  \mathcal{L}(\myedge{e}) 
\label{def-hcl-L}
\end{equation}
where $\mathsf{E} = \{ \myedge{e} \}$ is the set of edges of the lattice.
The edge function $\mathcal{L}(\myedge{e})$ depends on variables 
defined on the set of nodes $\mathsf{V} = \{ \mynode{v} \}$ of the lattice, 
$ 
  \mathcal{L}(\myedge{e}) 
  = 
  \mathcal{L}\left( 
    \mynode{v}_{+}(\myedge{e}), \mynode{v}_{-}(\myedge{e}) 
  \right) 
$  
where $\mynode{v}_{\pm}(\myedge{e})$ are the two nodes connected by the edge 
$\myedge{e}$ (so, in fact, it is an `interaction' function), and is given by 
\begin{equation}
  \mathcal{L}(\myedge{e}) 
  = 
  \Gamma(\myedge{e}) 
  \ln\left| 
    \myvar{u}(\mynode{v}_{+}(\myedge{e})) 
    - 
    \myvar{u}(\mynode{v}_{-}(\myedge{e})) 
  \right| 
\label{def-hcl-Le}
\end{equation} 
where  
$\Gamma(\myedge{e})$ are constants that depend only on the direction of the 
edge $\myedge{e}$, 
$\Gamma(\myedge{e}) \in \{ \Gamma_{1}, \Gamma_{2}, \Gamma_{3} \}$ 
(see equation \eref{def-gamma} below) and are subjected to the restriction, 
which appears in \cite{A01,BS02},  
\begin{equation} 
  \sum_{\myedge{e} \; (\mynode{v} \in \myedge{e}) } 
  \Gamma(\myedge{e}) 
  = 0,
  \qquad
  \mynode{v} \in \mathsf{V} 
\label{restr-ge}
\end{equation}
(we discuss this restriction in the Conclusion).
Here, $\myvar{u}=\myvar{u}(\mynode{v})$ is a function that should be found 
from the `variational' equations 
\begin{equation}
  \left. \partial \mathcal{S} \right/ \partial \myvar{u}(\mynode{v}) = 0, 
  \qquad
  \mynode{v} \in \mathsf{V}. 
\label{eq-variation}
\end{equation}

\begin{figure}%
\begin{center}
\includegraphics[width=\textwidth]{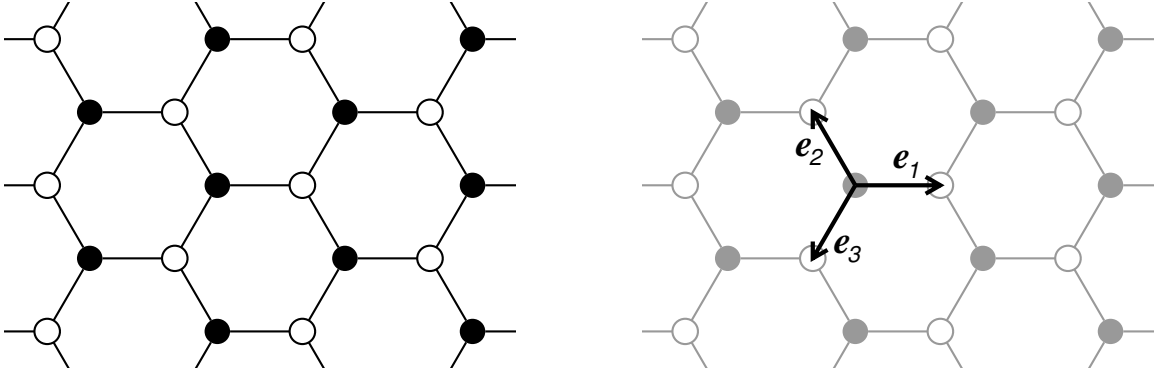}%
\end{center}
\caption{Bipartition of the HL and base vectors.}%
\label{fig-1}%
\end{figure}

In what follows, we extensively use the fact that the HL is a bipartite graph. 
So, from the beginning, we consider its set of vertices $\mathsf{V}$ as a sum 
of two subsets, which we call `positive' and `negative', 
$ \mathsf{V} = \mathsf{V}^{+} \cup \mathsf{V}^{-} $ 
(in figure \ref{fig-1}, the vertices that belong to $\mathsf{V}^{+}$ are shown 
by black circles and the vertices that belong to $\mathsf{V}^{-}$ are shown by 
white ones). Thus, the maps $\mynode{v}_{\pm}(\myedge{e})$ are maps to 
$\mathsf{V}^{\pm}$, 
$\mynode{v}_{\pm}: \mathsf{E} \to \mathsf{V}^{\pm}$.

Hereafter, instead of the $\myedge{e}$-$\mynode{v}$ notation, 
we use the vector one. To this end we introduce coplanar vectors 
$\vec{e}_{1}$, $\vec{e}_{2}$ and $\vec{e}_{3}$ related by 
\begin{equation}
  \sum_{i=1}^{3} \vec{e}_{i} = \vec{0} 
\label{e-condition}
\end{equation}
(see figure \ref{fig-1}), 
the set of lattice vectors $\Lambda$ (positions of the nodes of the HL), 
\begin{equation}
  \Lambda 
  = 
  \left\{\left. 
    \vec{n} = \sum_{i=1}^{3} n_{i} \vec{e}_{i}, 
    \quad
    n_{i} \in \mathbb{Z}
    \quad \right| \quad
    \sum_{i=1}^{3} n_{i} \ne 2 \, \mathop{\mbox{mod}} \, 3
  \right\}, 
\end{equation}
which can be decomposed as 
\begin{equation}
 \Lambda = \Lambda^{+} \cup \Lambda^{-} 
\end{equation} 
with 
\begin{equation}
  \begin{array}{l}
  \Lambda^{+} 
  = 
  \left\{ \left. 
    \vec{n} = \sum\limits_{i=1}^{3} n_{i} \vec{e}_{i}, 
    \quad
    n_{i} \in \mathbb{Z}
    \quad \right| \quad
    \sum\limits_{i=1}^{3} n_{i} = 0 \, \mathop{\mbox{mod}} \, 3 \; 
  \right\}, 
  \\[4mm] 
  \Lambda^{-} 
  = 
  \left\{ \left. 
    \vec{n} = \sum\limits_{i=1}^{3} n_{i} \vec{e}_{i}, 
    \quad
    n_{i} \in \mathbb{Z}
    \quad \right| \quad
    \sum\limits_{i=1}^{3} n_{i} = 1 \, \mathop{\mbox{mod}} \, 3 \; 
  \right\} 
  \end{array}
\end{equation}
(the vertex whose position is determined by $\vec{n} \in \Lambda^{\pm}$ 
belongs to $\mathsf{V}^{\pm}$) and write $\myvar{u}(\vec{n})$ instead of 
$\myvar{u}(\mynode{v})$.
It should be noted that, first, the usage of the three integer coordinates 
$n_{i}$ ($i=1,2,3$) does not mean that we are passing to the cubic lattice 
$\mathbb{Z}^{3}$ and, secondly, that
one should be careful and not forget the fact that the 
decomposition $\vec{n} = \sum_{i=1}^{3} n_{i}\vec{e}_{i}$ is not unique: 
triples $\left( n_{1}, n_{2}, n_{3} \right)$ and 
$\left( n_{1}+N, n_{2}+N, n_{3}+N \right)$ with integer non-zero $N$ define 
the same vector $\vec{n}$.

In vector terms, the action $\mathcal{S}$ \eref{def-hcl-L} can be presented as 
\begin{eqnarray}
  \mathcal{S}  
  & = & 
  \sum\limits_{\vec{n} \in \Lambda^{+}} 
  \sum\limits_{i=1}^{3} 
    \Gamma_{i} \,
    \ln \left| 
           \myvar[1]{u}(\vec{n}) - \myvar[2]{u}(\vec{n} + \vec{e}_{i}) 
        \right|
\\
  & = &
  \sum\limits_{\vec{n} \in \Lambda^{-}} 
  \sum\limits_{i=1}^{3} 
    \Gamma_{i} \, 
    \ln \left| 
           \myvar[2]{u}(\vec{n}) - \myvar[1]{u}(\vec{n} - \vec{e}_{i}) 
        \right|
\end{eqnarray}
where we use, instead of $\Gamma(\myedge{e})$, constants $\Gamma_{i}$, 
\begin{equation}
  \Gamma(\myedge{e}) = \Gamma_{i} 
  \quad \mbox{if} \quad 
  \myedge{e} \, \| \, \vec{e}_{i} 
  \quad
  (i=1,2,3), 
\label{def-gamma} 
\end{equation}
(where $\|$ stands for ``parallel'') subjected to the restriction 
\eref{restr-ge}, 
\begin{equation}
    \sum\limits_{i=1}^{3} \Gamma_{i} = 0.
\label{restr-gamma}
\end{equation}
The `variational' equations \eref{eq-variation} can now be written as 
\begin{eqnarray}
  &&
  \sum_{i=1}^{3} 
    \frac{ \Gamma_{i} }
         { \myvar[1]{u}(\vec{n}) - \myvar[2]{u}(\vec{n} + \vec{e}_{i}) }
  = 0 
  \qquad (\vec{n} \in \Lambda^{+}), 
\label{eq-hcl-p}
\\
  &&
  \sum_{i=1}^{3} 
    \frac{ \Gamma_{i} }
         { \myvar[2]{u}(\vec{n}) - \myvar[1]{u}(\vec{n} - \vec{e}_{i}) }
  = 0 
  \qquad (\vec{n} \in \Lambda^{-}).  
\label{eq-hcl-n}
\end{eqnarray}
Namely these equations are the main object of our study.

%%%%%%%%%%%%%%%%%%%%%%%%%%%%%%%%%%%%%%%%%%%%%%%%%%%%%%%%%%%%%%%%%%%%%%%%%%%%%%%
\subsection{Triangular lattice. \label{sec-def-t}}
%%%%%%%%%%%%%%%%%%%%%%%%%%%%%%%%%%%%%%%%%%%%%%%%%%%%%%%%%%%%%%%%%%%%%%%%%%%%%%%

It is straightforward to verify that any solution of 
\eref{eq-hcl-p} and \eref{eq-hcl-n} is, at the same time, a solution for the 
field equations of the model similar to \eref{def-hcl-L} and \eref{def-hcl-Le} 
but defined on the TL which is a sublattice 
of the HL discussed previously. 

\begin{figure}%
\begin{center}
\includegraphics[width=\textwidth]{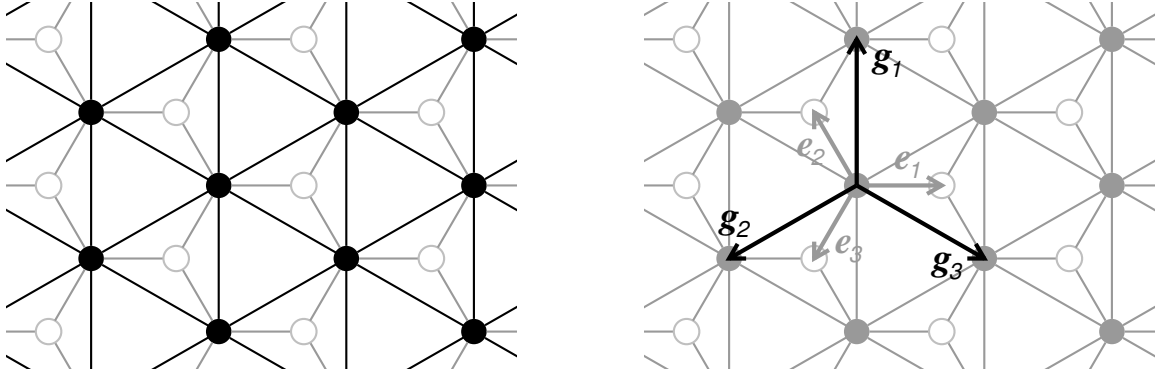}%
\end{center}
\caption{Honeycomb and triangular lattices}%
\label{fig-2}%
\end{figure}

Indeed, one can easily derive from \eref{eq-hcl-p} and \eref{eq-hcl-n} explicit 
expressions for $\myvar{u}(\vec{n})$ in terms of the values of $\myvar{u}$ 
at the adjacent points. In other words, we can eliminate, for example, the 
`negative' vertices,
\begin{equation}
  \myvar[2]{u}(\vec{n}) = 
  - 
  \frac{ 
    \sum_{i=1}^{3} 
    \Gamma_{i} \; 
    \myvar[1]{u}(\vec{n} - \vec{e}_{i-1}) 
    \myvar[1]{u}(\vec{n} - \vec{e}_{i+1}) 
  }
  { 
    \sum_{i=1}^{3} 
    \Gamma_{i} \; 
    \myvar[1]{u}(\vec{n} - \vec{e}_{i}) 
  }
  \qquad 
  (\vec{n} \in \Lambda^{-}).
\label{eq-tri-a}
\end{equation} 
In this  equation, as well as in the rest of the paper, 
we use the following convention: all arithmetic operations with 
$\vec{e}$- and $\Gamma$-indices are understood modulo $3$,
\begin{equation}
  \vec{e}_{i \pm 3} = \vec{e}_{i}, 
  \qquad 
  \Gamma_{i \pm 3} = \Gamma_{i}
  \qquad
  (i=1,2,3).
\label{def-indices}
\end{equation}
Equations \eref{eq-tri-a} and \eref{restr-gamma} lead to 
\begin{eqnarray}
  \frac{ \Gamma_{i} }
       { \myvar[1]{u}(\vec{n}) - \myvar[2]{u}(\vec{n} + \vec{e}_{i}) } 
  & = & 
  -   
  \frac{ \Gamma_{i-1} }
       { \myvar[1]{u}(\vec{n}) 
         - \myvar[1]{u}(\vec{n} + \vec{e}_{i} - \vec{e}_{i+1}) } 
\nonumber 
\\&&
  -   
  \frac{ \Gamma_{i+1} }
       { \myvar[1]{u}(\vec{n}) 
         - \myvar[1]{u}(\vec{n} + \vec{e}_{i} - \vec{e}_{i-1}) } 
  \qquad (\vec{n} \in \Lambda^{+}) 
\end{eqnarray}
Now, equation \eref{eq-hcl-p} implies 
\begin{equation}
  \sum_{i=1}^{3}
    \Gamma_{i} \left[ 
    \frac{ 1 }{ \myvar[1]{u}(\vec{n}) - \myvar[1]{u}(\vec{n} + \vec{g}_{i}) } 
    +
    \frac{ 1 }{ \myvar[1]{u}(\vec{n}) - \myvar[1]{u}(\vec{n} - \vec{g}_{i}) }
    \right] 
  = 
  0
  \qquad  (\vec{n} \in \Lambda^{+}) 
\label{eq-tri-p}
\end{equation}
where 
\begin{equation}
  \vec{g}_{1} = \vec{e}_{2} - \vec{e}_{3}, \qquad
  \vec{g}_{2} = \vec{e}_{3} - \vec{e}_{1}, \qquad
  \vec{g}_{3} = \vec{e}_{1} - \vec{e}_{2} 
\end{equation}
or 
\begin{equation}
  \vec{g}_{i} = \vec{e}_{i+1} - \vec{e}_{i-1}. 
\end{equation}
It is easy to see that these equations have the form 
$ \partial \mathcal{S}_{\mytriangle} 
  /
  \partial \myvar[1]{u}(\vec{n}) 
  = 0
$
with 
\begin{equation}
  \mathcal{S}_{\mytriangle}
  = 
  \sum_{\vec{n} \in \Lambda^{+}} 
  \sum_{i=1}^{3} 
    \Gamma_{i} \, 
    \ln \left| 
      \myvar[1]{u}(\vec{n}) - \myvar[1]{u}(\vec{n} + \vec{g}_{i}) 
    \right|. 
\label{def-L-tri}
\end{equation}
Thus, solutions for \eref{eq-hcl-p} and \eref{eq-hcl-n} solve also the field 
equations for the model similar to \eref{def-hcl-L} and \eref{def-hcl-Le}, 
with the same interaction, only, this time, along the edges of the TL 
$\Lambda^{+}$ (it is clear that one can repeat the same procedure to 
arrive at the lattice $\Lambda^{-}$).

%%%%%%%%%%%%%%%%%%%%%%%%%%%%%%%%%%%%%%%%%%%%%%%%%%%%%%%%%%%%%%%%%%%%%%%%%%%%%%%
\subsection{Cross-ratio system. \label{sec-crs}}
%%%%%%%%%%%%%%%%%%%%%%%%%%%%%%%%%%%%%%%%%%%%%%%%%%%%%%%%%%%%%%%%%%%%%%%%%%%%%%%

In this section we discuss the results of \cite{BS02} and their possible 
applications to the problem we are going to solve. 
The authors of \cite{BS02} developed a general construction that enables to 
transform a problem on a graph to the system of quad-equations. 
In particular, they considered a model 
(which they call ``additive rational Toda system'') 
which is a generalized version of the model discussed in this paper. 
Briefly, their scheme for the HL can described as follows: we extend our 
lattice by adding the vertices corresponding to the faces of the HL 
and consider equations we want to solve on 
$\Lambda^{ext} = \Lambda \cup \Lambda^{\circ}$, where 
\begin{equation}
  \Lambda^{\circ} 
  = 
  \left\{ \left. 
    \vec{n} = \sum\limits_{i=1}^{3} n_{i} \vec{e}_{i}, 
    \quad
    n_{i} \in \mathbb{Z}
    \quad \right| \quad
    \sum\limits_{i=1}^{3} n_{i} = 2 \mathop{\mbox{mod}} 3 \; 
  \right\}.
\end{equation}
It has been shown that, if one introduce $\myvar[1]{u}(\vec{n})$ with 
$\vec{n} \in \Lambda^{\circ}$ by the equations 
\begin{equation}
  \frac{ \Gamma_{i} }
       { \myvar[1]{u}(\vec{n}) - \myvar[2]{u}(\vec{n} + \vec{e}_{i}) } 
  = 
  \frac{ \gamma_{i+1} }
       { \myvar[1]{u}(\vec{n}) - \myvar[3]{u}(\vec{n} - \vec{e}_{i+1}) } 
  - 
  \frac{ \gamma_{i-1} }
       { \myvar[1]{u}(\vec{n}) - \myvar[3]{u}(\vec{n} - \vec{e}_{i-1}) } 
  \qquad  (\vec{n} \in \Lambda^{+}) 
\end{equation}
with 
\begin{equation}
  \Gamma_{i} = \gamma_{i+1} - \gamma_{i-1} 
  \qquad 
  (\gamma_{i \pm 3} = \gamma_{i}), 
\label{def-crs-gamma}
\end{equation}
then $\myvar{u}(\vec{n})$ is a solution for \eref{eq-hcl-p} provided 
it obey the so-called cross-ratio system on $\Lambda^{ext}$, 
\begin{eqnarray}
  0 
  & = & 
  \gamma_{i+1} 
  \left[ \myvar{u}(\vec{n} + \vec{e}_{i}) 
       - \myvar{u}(\vec{n} - \vec{e}_{i+1}) \right] 
  \left[ \myvar{u}(\vec{n}) 
       - \myvar{u}(\vec{n} - \vec{e}_{i-1}) \right] 
\nonumber \\
  & - &
  \gamma_{i-1} 
  \left[ \myvar{u}(\vec{n} + \vec{e}_{i}) 
       - \myvar{u}(\vec{n} - \vec{e}_{i-1}) \right] 
  \left[ \myvar{u}(\vec{n}) 
       - \myvar{u}(\vec{n} - \vec{e}_{i+1}) \right] 
  \qquad  
  (\vec{n} \in \Lambda^{+}). 
\end{eqnarray}
In a similar way, it can be shown that the system 
\begin{eqnarray}
  0 
  & = & 
  \gamma_{i+1} 
  \left[ \myvar{u}(\vec{n} - \vec{e}_{i}) 
       - \myvar{u}(\vec{n} + \vec{e}_{i+1}) \right] 
  \left[ \myvar{u}(\vec{n}) 
       - \myvar{u}(\vec{n} + \vec{e}_{i-1}) \right] 
\nonumber \\
  & - &
  \gamma_{i-1} 
  \left[ \myvar{u}(\vec{n} - \vec{e}_{i}) 
       - \myvar{u}(\vec{n} + \vec{e}_{i-1}) \right] 
  \left[ \myvar{u}(\vec{n}) 
       - \myvar{u}(\vec{n} + \vec{e}_{i+1}) \right] 
  \qquad  
  (\vec{n} \in \Lambda^{-}). 
\end{eqnarray}
implies \eref{eq-hcl-n}.
In both cases, one arrives at the set of the cross-ratio equations
\begin{eqnarray}
  \mathfrak{q}_{jk}(\vec{n})
  & := & 
  \gamma_{j} 
  \left[ \myvar{u}(\vec{n}) 
       - \myvar{u}(\vec{n} + \vec{e}_{k}) \right] 
  \left[ \myvar{u}(\vec{n} + \vec{e}_{j} + \vec{e}_{k}) 
       - \myvar{u}(\vec{n} + \vec{e}_{j}) \right] 
\nonumber \\
  & - &
  \gamma_{k} 
  \left[ \myvar{u}(\vec{n}) 
       - \myvar{u}(\vec{n} + \vec{e}_{j}) \right] 
  \left[ \myvar{u}(\vec{n} + \vec{e}_{j} + \vec{e}_{k}) 
       - \myvar{u}(\vec{n} + \vec{e}_{k}) \right] 
  = 
  0
\label{eq-cross-ratio}
\end{eqnarray}
where $(j,k) \in \{ (1,2), (1,3), (2,3) \}$. 
Thus, one may try to use the already known solutions for cross-ratio equation 
to obtain the ones for \eref{eq-hcl-p} and \eref{eq-hcl-n}. 
However, it turns out to be not a trivial problem. 
The case is that the cross-ratio system of \cite{BS02} is not the same that 
the cross-ratio equation of, for example, \cite{NQC83,NC95} or 
\cite{NAH09,HZ09}. The latter, is usually understood as an equation on 
$\mathbb{Z}^{2}$ whereas here, following the construction 
of \cite{BS02}, we have three types of quadrilaterals and hence three 
equations with different labelling (in the terminology of \cite{BS02}). 
Thus, to use the results of %, for example, 
\cite{NQC83,NC95,NAH09,HZ09} one has either 
to find common solutions for the all three cross-ratio equations or to 
find out how to `glue together' solutions for equations with different 
parameters. 
%(say, by using different solutions for the sublattices 
%$\Lambda^{+}$ and $\Lambda^{-}$). 
This is an interesting problem. which, however, is outside the scope of 
this paper. 
Moreover, in what follows we do not exploit the quad-equations approach and 
use, instead of the cross-ratio equation, another well-known system.

%%%%%%%%%%%%%%%%%%%%%%%%%%%%%%%%%%%%%%%%%%%%%%%%%%%%%%%%%%%%%%%%%%%%%%%%%%%%%%%%
\section{Bilinearization of the field equations. \label{sec-bilin}}
%%%%%%%%%%%%%%%%%%%%%%%%%%%%%%%%%%%%%%%%%%%%%%%%%%%%%%%%%%%%%%%%%%%%%%%%%%%%%%%%

In this section we present the main result of this paper. 
We bilinearize the field equations \eref{eq-hcl-p} and \eref{eq-hcl-n} and 
demonstrate the relationships of the resulting system with the already known 
integrable models. 
The procedure that we use is, for the most part, rather standard. However, there 
are few non-trivial moments that need some additional comments. 
We formulate some of the key statements `as is', without preliminary motivation, 
hoping that a reader will find the answers to possible questions in the 
following proofs and discussion.

We study a \emph{two}-dimensional lattice. However, we have the \emph{three} 
type of translations that correspond to the three vectors $\vec{e}_{i}$. 
Of course, one can introduce a basis consisting of two vectors and proceed in 
the standard manner,  
but, in our opinion, such approach has its disadvantages, most of which 
stem from the fact that it disregards the symmetry of the problem. 
Here we use another scheme. 
We consider our equations as a three-dimensional problem, bearing in mind 
that at some stage we have to take into account the fact that 
$\sum_{i=1}^{3}\vec{e}_{i}=\vec{0}$. 

The `three-dimensionality' of the equations that we want to solve does not 
provide insuperable problems: there are some already known integrable systems 
in dimensions higher than two (see, for example, \cite{ABS12,FNR15} for the 
lists of integrable discrete three-dimensional equations) and in what follows 
we use one of them. The second moment, stemming from the necessity to ensure 
the triviality of the superposition of the translations corresponding to 
$\sum_{i=1}^{3}\vec{e}_{i}$, turns out to be more embarrassing: 
a straightforward application of corresponding restrictions can drastically 
narrow the family of available solutions (we return to this question in the 
next sections when discussing the specific solutions). 
To overcome these difficulties, we `replace' the vectors $\vec{e}_{i}$ 
with \emph{arbitrary} vectors $\vec{\alpha}_{i}$, 
\begin{equation}
  \vec{e}_{i} \to \vec{\alpha}_{i} - \vec{\alpha}_{*}, 
\qquad 
  \vec{\alpha}_{*} 
  = 
  \frac{1}{3} \sum_{i=1}^{3} \vec{\alpha}_{i} 
\label{eq-e-aa}
\end{equation}
which automatically takes into account \eref{e-condition}.
In other words, we will consider, instead of $\myvar{u}(\vec{n})$, 
$\vec{n} = \sum\nolimits_{i=1}^{3} n_{i} \vec{e}_{i}$, 
functions of 
$
  \vec{\nu} 
  = 
  \sum\nolimits_{i=1}^{3} 
  n_{i} \left( \vec{\alpha}_{i} - \vec{\alpha}_{*}\right)
$.

Then, we make the substitution
\begin{equation}
  \myvar{u}(\vec{n}) 
  = 
  \left\{ 
  \begin{array}{ccl}
  q(\vec{\nu}) 
  & \qquad & 
  (\vec{n} \in \Lambda^{+}) 
  \\[2mm]
  \displaystyle 
  - 1 / r(\vec{\nu} - 2 \vec{\alpha}_{*}) 
  &&
  (\vec{n} \in \Lambda^{-}) 
  \end{array}
  \right.
\label{def-u-qr}
\end{equation}
and introduce the triplet of the tau-functions $(\sigma, \tau, \rho)$ by 
\begin{equation}
  q = \frac{ \sigma }{ \tau }, 
  \qquad 
  r = \frac{ \rho }{ \tau }. 
\label{def-tau}
\end{equation}
Finally, it can be shown that this construction leads to solutions for our 
equations \eref{eq-hcl-p} and \eref{eq-hcl-n} provided $\sigma$, $\tau$, and 
$\rho$ solve the following bilinear system: 
\begin{eqnarray}
  \sigma(\vec{\nu}+\vec{\alpha}) \tau(\vec{\nu}+\vec{\beta}) 
  - 
  \tau(\vec{\nu}+\vec{\alpha}) \sigma(\vec{\nu}+\vec{\beta}) 
  & = & 
  \myconst{a}{\vec{\alpha},\vec{\beta}} \, 
  \tau(\vec{\nu}) \sigma(\vec{\nu}+\vec{\alpha}+\vec{\beta}), 
\label{eq-bist}
\\
  \tau(\vec{\nu}+\vec{\alpha}) \rho(\vec{\nu}+\vec{\beta}) 
  - 
  \rho(\vec{\nu}+\vec{\alpha}) \tau(\vec{\nu}+\vec{\beta}) 
  & = & 
  \myconst{a}{\vec{\alpha},\vec{\beta}} \, 
  \rho(\vec{\nu}) \tau(\vec{\nu}+\vec{\alpha}+\vec{\beta}), 
\label{eq-bitr}
\\
  \tau(\vec{\nu}) \tau(\vec{\nu}+\vec{\alpha}+\vec{\beta}) 
  +
  \rho(\vec{\nu}) \sigma(\vec{\nu}+\vec{\alpha}+\vec{\beta}) 
  & = & 
  \myconst{b}{\vec{\alpha},\vec{\beta}} \, 
  \tau(\vec{\nu}+\vec{\alpha}) \tau(\vec{\nu}+\vec{\beta}) 
\label{eq-alh}
\end{eqnarray}
where skew-symmetric constants $\myconst{a}{\vec{\alpha},\vec{\beta}}$ 
and symmetric constants $\myconst{b}{\vec{\alpha},\vec{\beta}}$ 
are related to $\Gamma_{i}$ by 
\begin{equation}
  \Gamma_{i}
  = 
  \myconst{a}{\vec{\alpha}_{i+1},\vec{\alpha}_{i-1}}
  \myconst{b}{\vec{\alpha}_{i+1},\vec{\alpha}_{i-1}}.
\label{def-gamma-ab}
\end{equation}
Here (and in what follows) we use the convention similar to \eref{def-indices}, 
\begin{equation}
  \vec{\alpha}_{i \pm 3} = \vec{\alpha}_{i} 
  \qquad
  (i=1,2,3).
\end{equation}

The main differences between the original system \eref{eq-hcl-p}, 
\eref{eq-hcl-n} and system \eref{eq-bist}--\eref{eq-alh} (except that the 
latter is bilinear) are the following. 
First, whereas equations \eref{eq-hcl-p} and \eref{eq-hcl-n} are defined on the 
sublattices of the two-dimensional HL ($\Lambda^{+}$ and $\Lambda^{-}$), 
in \eref{eq-bist}--\eref{eq-alh} one has to deal with the 
lattice which is $\mathbb{Z}^{3}$ 
(the parameters $\vec\alpha$ and $\vec\beta$ belong to 
$\{ \vec{\alpha}_{1}, \vec{\alpha}_{2}, \vec{\alpha}_{3} \}$). 
Secondly, instead of two subsystems (\eref{eq-hcl-p} for $\Lambda^{+}$ and 
\eref{eq-hcl-n} for $\Lambda^{-}$) we consider \emph{all} equations of 
\eref{eq-bist}--\eref{eq-alh} as defined on \emph{all} points of 
$\mathbb{Z}^{3}$.
Thus, whereas the tau-functions $\sigma$ and $\rho$ were introduced in 
\eref{def-tau} for different sublattices of the HL 
($\Lambda^{+}$ and $\Lambda^{-}$ correspondingly), in the framework 
of \eref{eq-bist}--\eref{eq-alh} they are parts of the triplet 
$\{ \sigma, \tau, \rho \}$ which is associated with \emph{each} point 
of $\mathbb{Z}^{3}$. In other words, we have passed from the bipartite system 
\eref{eq-hcl-p} and \eref{eq-hcl-n}, which reflects the geometry of the HL, 
to a translationally-invariant system \eref{eq-bist}--\eref{eq-alh} which is 
easier to handle, for example, when one is looking for solutions. 

Sometimes, to avoid writing separate formulae for $\Lambda^{+}$ and 
$\Lambda^{-}$, we will take into account the summand $2\vec{\alpha}_{*}$ in 
\eref{def-u-qr} by introducing $\vec{\nu}_{\pm}$, 
\begin{equation}
  \begin{array}{lcl}
  \vec{\nu}_{+}(\vec{n}) = \vec{\nu}(\vec{n})
  & \qquad & 
  (\vec{n} \in \Lambda^{+}) 
  \\ 
  \vec{\nu}_{-}(\vec{n}) 
  = 
  \vec{\nu}(\vec{n}) - 2\vec{\alpha}_{*}
  &&
  (\vec{n} \in \Lambda^{-}) 
  \end{array}
\end{equation}
that can be presented as 
\begin{equation}
  \vec{\nu}_{\pm}\left(\sum_{i=1}^{3} n_{i}\vec{e}_{i} \right) 
  = 
  \sum_{i=1}^{3} 
  \left[ n_{i} - \mathcal{N}\left(n_{1},n_{2},n_{3}\right) \right] 
  \vec{\alpha}_{i}
\label{def-nu-pm}
\end{equation}
with 
\begin{equation}
  \mathcal{N}\left(n_{1},n_{2},n_{3}\right) 
  = 
  \left\{ 
  \begin{array}{lcl}
  \frac{1}{3} \sum_{i=1}^{3} n_{i} 
  & \quad & 
  ( \sum_{i=1}^{3} n_{i} = 0 \, \mathop{\mbox{mod}} 3 )
  \\[2mm]
  \frac{1}{3} \left( \sum_{i=1}^{3} n_{i} + 2 \right) 
  & \quad & 
  ( \sum_{i=1}^{3} n_{i} = 1 \, \mathop{\mbox{mod}} 3 ) 
  \end{array}
  \right.
\end{equation}
or, alternatively,
\begin{equation}
  \mathcal{N}\left(n_{1},n_{2},n_{3}\right) 
  = 
  \left\lfloor 
    \frac{1}{3} \left( \sum_{i=1}^{3} n_{i} + 2 \right) 
  \right\rfloor 
\label{def-N-cal}
\end{equation}
where $\lfloor ... \rfloor$ stands for the floor function (integer part): 
for any integer $N$ and $0 \le \delta < 1$, 
$\lfloor N+\delta \rfloor = N$. 
Note that the differences 
$n_{i} - \mathcal{N}\left(n_{1},n_{2},n_{3}\right)$ do not depend on the 
decomposition of a lattice vector $\vec{n}$ into 
$\vec{n} = \sum_{i=1}^{3} n_{i}\vec{e}_{i}$.

To summarize, the main result of this paper can be presented as 

%%%%%%%%%%%%%%%%%%%%%%%%%%%%%%%%%%%%%%%%%%%%%%%%%%%%%%%%%%%%%%%%%%%%%%%%%%%%%%%%
\begin{proposition} \label{prop-bilin}
Any solution for the bilinear system \eref{eq-bist}--\eref{eq-alh} 
with the coefficients that satisfy the restriction \eref{def-gamma-ab} 
provide a solution for the nonlinear HL equations 
\eref{eq-hcl-p} and \eref{eq-hcl-n}, 
as well as for the nonlinear TL equations \eref{eq-tri-p}, 
which can be obtained by 
\begin{equation}
  \myvar{u}(\vec{n}) 
  = 
  \left\{ 
  \begin{array}{ccl}
  \sigma(\vec{\nu}_{+}(\vec{n})) / \tau(\vec{\nu}_{+}(\vec{n})) 
  & \qquad & 
  (\vec{n} \in \Lambda^{+}) 
  \\[4mm]
  - \tau(\vec{\nu}_{-}(\vec{n})) / \rho(\vec{\nu}_{-}(\vec{n})) 
  &&
  (\vec{n} \in \Lambda^{-}) 
  \end{array}
  \right.
\end{equation}
where vectors $\vec{\nu}_{\pm}(\vec{n})$ are defined in \eref{def-nu-pm}. 
\end{proposition}
%%%%%%%%%%%%%%%%%%%%%%%%%%%%%%%%%%%%%%%%%%%%%%%%%%%%%%%%%%%%%%%%%%%%%%%%%%%%%%%%

In the following subsections we prove this proposition 
by demonstrating how equations \eref{eq-bist}--\eref{eq-alh} 
`help' us to solve the field equations \eref{eq-hcl-p}, \eref{eq-hcl-n} 
and \eref{eq-tri-p}. 

%%%%%%%%%%%%%%%%%%%%%%%%%%%%%%%%%%%%%%%%%%%%%%%%%%%%%%%%%%%%%%%%%%%%%%%%%%%%%%%%
\subsection{Solving equations \eref{eq-hcl-p}.}

Let us consider a `positive' vertex, $\vec{n} \in \Lambda^{+}$. 
It follows from definition \eref{def-u-qr} that 
\begin{equation}
  \myvar[1]{u}(\vec{n}) = q(\vec{\nu}), 
  \qquad
  \myvar[2]{u}(\vec{n} + \vec{e}_{i}) 
  = 
  - 1 / r(\vec{\nu} - \vec{\alpha}_{i+1} - \vec{\alpha}_{i-1}) 
\end{equation}
which leads to 
\begin{equation}
  \myvar[1]{u}(\vec{n}) 
  - 
  \myvar[2]{u}(\vec{n} + \vec{e}_{i}) 
  = 
  \frac{ 
    \tau(\vec{\nu} - \vec{\alpha}_{i+1} - \vec{\alpha}_{i-1}) 
    \tau(\vec{\nu}) 
    + 
    \rho(\vec{\nu} - \vec{\alpha}_{i+1} - \vec{\alpha}_{i-1}) 
    \sigma(\vec{\nu}) 
    }
    {
    \rho(\vec{\nu} - \vec{\alpha}_{i+1} - \vec{\alpha}_{i-1}) 
    \tau(\vec{\nu}) 
    }.
\end{equation}
Equation \eref{eq-alh} factorizes the numerator, 
\begin{equation}
  \myvar[1]{u}(\vec{n}) 
  - 
  \myvar[2]{u}(\vec{n} + \vec{e}_{i}) 
  = 
  \myconst{b}{\vec{\alpha}_{i+1},\vec{\alpha}_{i-1}} 
  \frac{ 
    \tau(\vec{\nu} - \vec{\alpha}_{i+1}) 
    \tau(\vec{\nu} - \vec{\alpha}_{i-1}) 
    }
    {
    \rho(\vec{\nu} - \vec{\alpha}_{i+1} - \vec{\alpha}_{i-1}) 
    \tau(\vec{\nu}) 
    },
\end{equation}
which leads, together with \eref{def-gamma-ab} and \eref{eq-bitr}, to 
\begin{equation}
  \frac{ \Gamma_{i} } 
       { \myvar[1]{u}(\vec{n}) - \myvar[2]{u}(\vec{n} + \vec{e}_{i}) }
  = 
  r(\vec{\nu} - \vec{\alpha}_{i+1}) 
  - 
  r(\vec{\nu} - \vec{\alpha}_{i-1}). 
\end{equation}
Now, it is clear that the summation over $i=1,2,3$ yields \eref{eq-hcl-p}.

This proves that any solution of \eref{eq-bist}--\eref{def-gamma-ab} provides 
a solution for \eref{eq-hcl-p}.

%%%%%%%%%%%%%%%%%%%%%%%%%%%%%%%%%%%%%%%%%%%%%%%%%%%%%%%%%%%%%%%%%%%%%%%%%%%%%%%%
\subsection{Solving equations \eref{eq-hcl-n}.}

In a similar way one can demonstrate that system \eref{eq-bist}--\eref{eq-alh} 
yields solutions for \eref{eq-hcl-n}. 
For any $\vec{n} \in \Lambda^{-}$, 
\begin{equation}
  \myvar[2]{u}(\vec{n}) 
  = 
  - 1 / r(\vec{\nu}'), 
  \qquad
  \myvar[1]{u}(\vec{n} - \vec{e}_{i}) 
  =  
  q(\vec{\nu}' + \vec{\alpha}_{i+1} + \vec{\alpha}_{i-1}) 
  \qquad
  (\vec{\nu}' = \vec{\nu}_{-}(\vec{n})) 
\end{equation}
and 
\begin{equation}
  \myvar[2]{u}(\vec{n}) 
  - 
  \myvar[1]{u}(\vec{n} - \vec{e}_{i}) 
  = 
  - \frac{ 
    \tau(\vec{\nu}') 
    \tau(\vec{\nu}' + \vec{\alpha}_{i+1} + \vec{\alpha}_{i-1}) 
    +
    \rho(\vec{\nu}') 
    \sigma(\vec{\nu}' + \vec{\alpha}_{i+1} + \vec{\alpha}_{i-1}) 
    }
    {
    \rho(\vec{\nu}') 
    \tau(\vec{\nu}' + \vec{\alpha}_{i+1} + \vec{\alpha}_{i-1}) 
    }.
\end{equation}
After application of \eref{eq-alh} and \eref{eq-bitr} the above equation 
leads to 
\begin{equation}
  \myvar[2]{u}(\vec{n}) 
  - 
  \myvar[1]{u}(\vec{n} - \vec{e}_{i}) 
  = 
  - 
  \myconst{b}{\vec{\alpha}_{i+1},\vec{\alpha}_{i-1}} 
  \frac{ 
    \tau(\vec{\nu}' + \vec{\alpha}_{i+1}) 
    \tau(\vec{\nu}' + \vec{\alpha}_{i-1}) 
    }
    {
    \rho(\vec{\nu}') 
    \tau(\vec{\nu}' + \vec{\alpha}_{i+1} + \vec{\alpha}_{i-1}) 
    }
\end{equation}
and 
\begin{equation}
  \frac{ \Gamma_{i} } 
       { \myvar[2]{u}(\vec{n}) - \myvar[1]{u}(\vec{n} - \vec{e}_{i}) }
  = 
  r(\vec{\nu}' + \vec{\alpha}_{i+1}) 
  - 
  r(\vec{\nu}' + \vec{\alpha}_{i-1}) 
\end{equation}
which demonstrates that $\myvar[4]{u}(\vec{n})$ solves \eref{eq-hcl-n}, 
i.e. that any solution of \eref{eq-bist}--\eref{def-gamma-ab} provides a solution 
for \eref{eq-hcl-n}.

%%%%%%%%%%%%%%%%%%%%%%%%%%%%%%%%%%%%%%%%%%%%%%%%%%%%%%%%%%%%%%%%%%%%%%%%%%%%%%%%
\subsection{Solving equations \eref{eq-tri-p}.}

To conclude this section we give another proof of the fact that the 
proposed construction \eref{def-u-qr}--\eref{def-gamma-ab} provides 
solutions for equations \eref{eq-tri-p} for the TL.
 
For any $\vec{n} \in \Lambda^{+}$, 
\begin{equation}
  \myvar[1]{u}(\vec{n}) = q(\vec{\nu}), 
  \qquad
  \myvar[1]{u}(\vec{n} + \vec{g}_{i}) 
  =  
  q(\vec{\nu} + \vec{\alpha}_{i+1} - \vec{\alpha}_{i-1}) 
\end{equation}
and 
\begin{equation}
  \myvar[1]{u}(\vec{n}) - \myvar[1]{u}(\vec{n} + \vec{g}_{i})  
  = 
  \frac{ 
    \sigma(\vec{\nu}) 
    \tau(\vec{\nu} + \vec{\alpha}_{i+1} - \vec{\alpha}_{i-1}) 
    - 
    \tau(\vec{\nu}) 
    \sigma(\vec{\nu} + \vec{\alpha}_{i+1} - \vec{\alpha}_{i-1}) 
    }
    {
    \tau(\vec{\nu}) 
    \tau(\vec{\nu} + \vec{\alpha}_{i+1} - \vec{\alpha}_{i-1}) 
    }
\end{equation}
which, with the help of \eref{eq-bist}, can be rewritten as 
\begin{equation}
  \myvar[1]{u}(\vec{n}) 
  - 
  \myvar[1]{u}(\vec{n} + \vec{g}_{i}) 
  = 
  - 
  \myconst{a}{\vec{\alpha}_{i+1},\vec{\alpha}_{i-1}} 
  \frac{ 
    \sigma(\vec{\nu} + \vec{\alpha}_{i+1}) 
    \tau(\vec{\nu} - \vec{\alpha}_{i-1}) 
    }
    {
    \tau(\vec{\nu}) 
    \tau(\vec{\nu} + \vec{\alpha}_{i+1} - \vec{\alpha}_{i-1}) 
    }.
\end{equation}
This, together with \eref{eq-alh}, leads to 
\begin{equation}
  \frac{ \Gamma_{i} } 
       { \myvar[1]{u}(\vec{n}) - \myvar[1]{u}(\vec{n} + \vec{g}_{i}) }
  = 
  - q^{-1}(\vec{\nu} + \vec{\alpha}_{i+1}) 
  - r(\vec{\nu} - \vec{\alpha}_{i-1}) 
\end{equation}
and then to 
\begin{equation}
  \frac{ \Gamma_{i} } 
       { \myvar[1]{u}(\vec{\nu}) - \myvar[1]{u}(\vec{n} + \vec{g}_{i}) }
  + 
  \frac{ \Gamma_{i} } 
       { \myvar[1]{u}(\vec{\nu}) - \myvar[1]{u}(\vec{n} - \vec{g}_{i}) }
  = 
  \myvar{w}_{i+1}(\vec{\nu}) 
  - 
  \myvar{w}_{i-1}(\vec{\nu}) 
\label{eq-tri-b}
\end{equation}
where 
\begin{equation}
  \myvar{w}_{i}(\vec{\nu}) 
  = 
  r(\vec{\nu} - \vec{\alpha}_{i})
  - q^{-1}(\vec{\nu} + \vec{\alpha}_{i}).
\end{equation}
Again, the structure of the summand in \eref{eq-tri-p} 
exposed in \eref{eq-tri-b} leads to the fact that the summation over $i$ 
produces zero result.
This proves that any solution of \eref{eq-bist}--\eref{def-gamma-ab} provides a 
solution for \eref{eq-tri-p}.

%%%%%%%%%%%%%%%%%%%%%%%%%%%%%%%%%%%%%%%%%%%%%%%%%%%%%%%%%%%%%%%%%%%%%%%%%%%%%%%%
\subsection{M\"obius invariance. \label{sec-mobius}}
%%%%%%%%%%%%%%%%%%%%%%%%%%%%%%%%%%%%%%%%%%%%%%%%%%%%%%%%%%%%%%%%%%%%%%%%%%%%%%%%

Another interesting fact that has not been mentioned yet, is the invariance of 
the field equations for HL or TL with respect to the M\"obius transformations: 

\begin{proposition} \label{prop-mobius}
If $\myvar{u}(\vec{n})$ solves \eref{eq-hcl-p} and \eref{eq-hcl-n}, 
or \eref{eq-tri-p}, so does 
\begin{equation}
  \frac{ a \, \myvar{u}(\vec{n}) + b } 
       { c \, \myvar{u}(\vec{n}) + d }
\end{equation}
with constant $a$, $b$, $c$ and $d$, $ad-bc \ne 0$.
\end{proposition} 

The proof of this statement is straightforward and is not presented here.

Thus, one can add three arbitrary constants to any solution presented in the  
following sections. 

%%%%%%%%%%%%%%%%%%%%%%%%%%%%%%%%%%%%%%%%%%%%%%%%%%%%%%%%%%%%%%%%%%%%%%%%%%%%%%%%
\section{Exact solutions for the HL. \label{sec-exact-h}}
%%%%%%%%%%%%%%%%%%%%%%%%%%%%%%%%%%%%%%%%%%%%%%%%%%%%%%%%%%%%%%%%%%%%%%%%%%%%%%%%

In this section we discuss system \eref{eq-bist}--\eref{eq-alh} and then 
present the two sets of exact solutions for the field 
equations \eref{eq-hcl-p} and \eref{eq-hcl-n} which are obtained by 
modification of the already known ones that have been derived for 
\eref{eq-bist}--\eref{eq-alh}. 

%%%%%%%%%%%%%%%%%%%%%%%%%%%%%%%%%%%%%%%%%%%%%%%%%%%%%%%%%%%%%%%%%%%%%%%%%%%%%%%% 
\subsection{Ablowitz-Ladik-Hirota system. \label{sec-alh}} 
%%%%%%%%%%%%%%%%%%%%%%%%%%%%%%%%%%%%%%%%%%%%%%%%%%%%%%%%%%%%%%%%%%%%%%%%%%%%%%%% 

Here, we collect some known facts about the system 
\eref{eq-bist}--\eref{eq-alh}, which we write now as
\begin{eqnarray}
  0 & = & 
    \myconst{a}{\alpha,\beta}
    \myvar{\tau} 
    \myShiftedB{\alpha\beta}{\myvar{\sigma}} 
  - \myShiftedB{\alpha}{\myvar{\sigma}} 
    \myShiftedB{\beta}{\myvar{\tau}} 
  + \myShiftedB{\alpha}{\myvar{\tau}} 
    \myShiftedB{\beta}{\myvar{\sigma}}, 
\label{alh-bist}
\\
  0 & = & 
    \myconst{a}{\alpha,\beta}
    \myvar{\rho} 
    \myShiftedB{\alpha\beta}{\myvar{\tau}} 
  - \myShiftedB{\alpha}{\myvar{\tau}} 
    \myShiftedB{\beta}{\myvar{\rho}} 
  + \myShiftedB{\alpha}{\myvar{\rho}} 
    \myShiftedB{\beta}{\myvar{\tau}}, 
\label{alh-bitr}
\\
  0 & = & 
    \myconst{b}{\alpha,\beta}
    \myShiftedB{\alpha}{\myvar{\tau}} 
    \myShiftedB{\beta}{\myvar{\tau}} 
  - \myvar{\tau} 
    \myShiftedB{\alpha\beta}{\myvar{\tau}} 
  - \myvar{\rho} 
    \myShiftedB{\alpha\beta}{\myvar{\sigma}} 
\label{alh-alh}
\end{eqnarray}%
where we use the `abstract shift' notation. 
Further, we recall that the shifts $\myShift{\alpha}$  
are, in our case, a way to write the translations 
$ \myShift\alpha: f(\vec{\nu}) \to f(\vec{\nu} + \vec{\alpha}) $ 
and identify the parameters $\alpha$ with the vectors $\vec{\alpha}_{i}$. 
However, now we want to present some simple, algebraic, consequences of 
system \eref{alh-bist}--\eref{alh-alh} which do not depend on the origin 
of these equations and the `inner structure' of the tau-functions. 
Thus, one can think of \eref{alh-bist}--\eref{alh-alh} as a system of 
difference (or functional) equations with arbitrary, save the consistency 
restriction that we discuss below (see \eref{alh-cab}), skew-symmetric functions 
$\myconst{a}{\alpha,\beta}$ and symmetric functions $\myconst{b}{\alpha,\beta}$ 
of arbitrary (scalar or vector) parameters $\alpha$ and $\beta$. 

It can be shown that an immediate consequence of, say, \eref{alh-bitr} is 
the fact that $\myvar{\tau}$ solves the famous Hirota bilinear difference 
equation (HBDE) \cite{H81}, also known as the discrete KP equation: 
\begin{equation}
  0 
  = 
    \myconst{a}{\alpha,\beta}
    \myShiftedB{\gamma}{\myvar{\tau}} 
    \myShiftedB{\alpha\beta}{\myvar{\tau}} 
  - \myconst{a}{\alpha,\gamma}
    \myShiftedB{\beta}{\myvar{\tau}} 
    \myShiftedB{\alpha\gamma}{\myvar{\tau}} 
  + \myconst{a}{\beta,\gamma}
    \myShiftedB{\alpha}{\myvar{\tau}} 
    \myShiftedB{\beta\gamma}{\myvar{\tau}} 
\label{alh-hbde}
\end{equation}
(we prove this statement in appendix \ref{app-hbde-1}). 
Moreover, it turns out that functions $\myvar{\sigma}$ and $\myvar{\rho}$ 
also solve \eref{alh-hbde}, 
\begin{equation}
  0 
  = 
    \myconst{a}{\alpha,\beta}
    \myShiftedB{\gamma}{\myvar{\omega}} 
    \myShiftedB{\alpha\beta}{\myvar{\omega}} 
  - \myconst{a}{\alpha,\gamma}
    \myShiftedB{\beta}{\myvar{\omega}} 
    \myShiftedB{\alpha\gamma}{\myvar{\omega}} 
  + \myconst{a}{\beta,\gamma}
    \myShiftedB{\alpha}{\myvar{\omega}} 
    \myShiftedB{\beta\gamma}{\myvar{\omega}}, 
  \quad
  \omega = \sigma, \rho 
\label{alh-hbde-sr}
\end{equation}
(see appendix \ref{app-hbde-1}). 
Thus, equation \eref{alh-bitr}, or \eref{alh-bist}, can be viewed as the 
linear problem (or the so-called Lax representation) for the HBDE 
(see, e.g., \cite{DJM82a,DJM82b,N97,WTS97,TSW97}). 
On the other hand, another consequences of 
equations \eref{alh-bist} and \eref{alh-bitr},  
\begin{eqnarray}
  0 & = & 
    \myconst{a}{\alpha,\beta}
    \myShiftedB{\gamma}{\myvar{\tau}} 
    \myShiftedB{\alpha\beta}{\myvar{\sigma}} 
  - \myconst{a}{\alpha,\gamma}
    \myShiftedB{\beta}{\myvar{\tau}} 
    \myShiftedB{\alpha\gamma}{\myvar{\sigma}} 
  + \myconst{a}{\beta,\gamma}
    \myShiftedB{\alpha}{\myvar{\tau}} 
    \myShiftedB{\beta\gamma}{\myvar{\sigma}}, 
\label{alh-bt-st}
\\
  0 & = & 
    \myconst{a}{\alpha,\beta}
    \myShiftedB{\gamma}{\myvar{\rho}} 
    \myShiftedB{\alpha\beta}{\myvar{\tau}} 
  - \myconst{a}{\alpha,\gamma}
    \myShiftedB{\beta}{\myvar{\rho}} 
    \myShiftedB{\alpha\gamma}{\myvar{\tau}} 
  + \myconst{a}{\beta,\gamma}
    \myShiftedB{\alpha}{\myvar{\rho}} 
    \myShiftedB{\beta\gamma}{\myvar{\tau}} 
\label{alh-bt-tr}
\end{eqnarray}
(see appendix \ref{app-hbde-2} for a proof), 
can be interpreted as describing the B\"acklund transformations 
\begin{equation}
  \mbox{BT}_{\mbox{\tiny HBDE}}: \qquad 
  \sigma 
  \stackrel{\eref{alh-bt-st}}{\longrightarrow} 
  \tau 
  \stackrel{\eref{alh-bt-tr}}{\longrightarrow}  
  \rho
\end{equation}
between different solutions for the HBDE 
(note that this chain can be continued in both directions,
$... \to \sigma \to \tau \to \rho \to ...$ via the extended version of 
\eref{alh-bist} and \eref{alh-bitr}). 
 
Clearly, one can derive a great number of identities for the functions that 
satisfy \eref{alh-bist} and \eref{alh-bitr}. Here, we write down only one 
example, 
\begin{equation}
    \myconst{A}{\alpha,\beta,\gamma}
    \myvar{\tau} 
    \myShiftedB{\alpha\beta\gamma}{\myvar{\sigma}} 
  = 
    \myconst{a}{\beta,\gamma}
    \myShiftedB{\alpha}{\myvar{\sigma}} 
    \myShiftedB{\beta\gamma}{\myvar{\tau}} 
  - \myconst{a}{\alpha,\gamma}
    \myShiftedB{\beta}{\myvar{\sigma}} 
    \myShiftedB{\alpha\gamma}{\myvar{\tau}} 
  + \myconst{a}{\alpha,\beta}
    \myShiftedB{\gamma}{\myvar{\sigma}} 
    \myShiftedB{\alpha\beta}{\myvar{\tau}} 
\label{alh-super-s}
\end{equation}
with 
$
  \myconst{A}{\alpha,\beta,\gamma}
  = 
  \myconst{a}{\alpha,\beta} 
  \myconst{a}{\alpha,\gamma} 
  \myconst{a}{\beta,\gamma}
$
which may be useful, if one wants to demonstrate the so-called 
three-dimensional consistency of \eref{alh-bist} and \eref{alh-bitr} 
(see appendix \ref{app-hbde-3}).

Till now, we have considered only the first two equations of the 
\eref{alh-bist}--\eref{alh-alh}. The last one can be viewed in the framework 
of the theory of the HBDE as a nonlinear restriction, which is compatible 
with \eref{alh-bist} and \eref{alh-bitr} provided the constants  
$\myconst{a}{\alpha,\beta}$ and $\myconst{b}{\alpha,\beta}$ met the following 
condition: 
\begin{equation}
    \myconst{a}{\alpha,\beta}
    \myconst{b}{\alpha,\beta}
  - \myconst{a}{\alpha,\gamma}
    \myconst{b}{\alpha,\gamma}
  + \myconst{a}{\beta,\gamma}
    \myconst{b}{\beta,\gamma} 
  = 0
\label{alh-cab}
\end{equation}
which is derived in appendix \ref{app-hbde-4}.

It turns out that the restricted system \eref{eq-bist}--\eref{eq-alh} is 
closely related to another integrable model, which is even `older' than 
the HBDE: equations \eref{eq-bist}--\eref{eq-alh} describe 
the so-called Miwa shifts of the Ablowitz-Ladik hierarchy (ALH) \cite{AL75}. 

Indeed, as is demonstrated in \ref{app-alh}, the functions 
\begin{equation}
  \myvar{Q} 
  = 
  \frac{ \myvar{E} }{ \myconst{b}{\kappa,\kappa} }
  \frac{ \myShifted{\kappa}{\myvar{\sigma}} }{ \myvar{\tau} }, 
\qquad
  \myvar{R} 
  = 
  \frac{ 1 }{ \myvar{E} } 
  \frac{ \myShift{\kappa}^{-1} \myvar{\rho} }{ \myvar{\tau} }
\label{alh-def-QR}
\end{equation}
where $\myvar{E}$ is defined by 
\begin{equation}
  \myShifted{\alpha}{\myvar{E}} 
  = 
  \frac{ 1 }{ \myconst{b}{\alpha,\kappa} } \; 
  \myvar{E} 
\label{alh-def-E}
\end{equation}
satisfy, for a fixed value of $\kappa$,
\begin{eqnarray}
    \myShift[E]{\alpha} \myvar{Q} 
  - \myvar{Q} 
  & = & 
    \xi_{\alpha} 
    \left[
      1 - \myvar{R} \left( \myShift[E]{\alpha} \myvar{Q} \right) 
    \right]
    \myShiftedB{\alpha}{\myvar{Q}} 
\label{eq-alh-1}
\\
    \myvar{R} 
  - \myShift[E]{\alpha} \myvar{R} 
  & = & 
    \xi_{\alpha} 
    \left[
      1 - \myvar{R} \left( \myShift[E]{\alpha} \myvar{Q} \right) 
    \right]
    \myShift{\kappa}^{-1} \myvar{R} 
\label{eq-alh-2}
\end{eqnarray}
where 
$\myShift[E]{\alpha} = \myShift{\alpha}\myShift{\kappa}^{-1}$ 
and 
$\xi_{\alpha} =  \myconst{a}{\alpha,\kappa}\myconst{b}{\alpha,\kappa}$.
Introducing the $n$-dependence by 
\begin{equation}
  Q_{n} = \myShift{\kappa}^{n} Q, 
  \qquad
  R_{n} = \myShift{\kappa}^{-n} R 
\end{equation}
one can rewrite \eref{eq-alh-1} and \eref{eq-alh-2} as 
\begin{eqnarray}
    \myShift[E]{\alpha} \myvar{Q}_{n} 
  - \myvar{Q}_{n} 
  & = & 
    \xi_{\alpha} 
    \left[
      1 - \myvar{R}_{n} \left( \myShift[E]{\alpha} \myvar{Q}_{n} \right) 
    \right]
    \myShift[E]{\alpha} \myvar{Q}_{n+1}, 
\label{eq-alh-3}
\\
    \myvar{R}_{n}  
  - \myShift[E]{\alpha} \myvar{R}_{n}  
  & = & 
    \xi_{\alpha} 
    \left[
      1 - \myvar{R}_{n} \left( \myShift[E]{\alpha} \myvar{Q}_{n} \right) 
    \right]
    \myvar{R}_{n-1}. 
\label{eq-alh-4}
\end{eqnarray}
These equations are nothing but the so-called functional representation of the 
positive flows of the ALH \cite{V98,V02}: 
if we consider $Q_{n}$ and $R_{n}$ as functions of an infinite 
number of variables, 
$Q_{n} = Q_{n}\left( z_{m} \right)_{m=1, ..., \infty}$, 
$R_{n} = R_{n}\left( z_{m} \right)_{m=1, ..., \infty}$ 
and identify the shifts $\myShift[E]\alpha$ with the Miwa shifts, 
\begin{equation}
  \myShift[E]\alpha 
  F_{n}\left( z_{m} \right)_{m=1, ..., \infty} 
  = 
  F_{n}\left( z_{m} + i \xi_{\alpha}^{m}/m \right)_{m=1, ..., \infty}, 
\end{equation}
then equations \eref{eq-alh-3} and \eref{eq-alh-4} can be viewed as an 
infinite set of differential equation, the simplest of which are given by  
\begin{eqnarray}
  i \left. \partial \myvar{Q}_{n} \right/ \partial z_{1}   
  & = & 
  \left[ 1 - \myvar{Q}_{n} \myvar{R}_{n} \right] \myvar{Q}_{n+1}, 
\\
  - i \left. \partial \myvar{R}_{n} \right/ \partial z_{1} 
  & = & 
  \left[ 1 - \myvar{Q}_{n} \myvar{R}_{n} \right] \myvar{R}_{n-1} 
\end{eqnarray}
(the complex version of the discrete nonlinear Schr\"odinger equation). 
This infinite set of differential equation is the positive part of the ALH. 
We do not discuss here the negative ALH, whose equations as well can be 
`derived' from \eref{alh-bist}--\eref{alh-alh}, referring to, for example, 
section 6 of \cite{V15} for details. What is important for our present study 
is that the ALH (and hence the system \eref{alh-bist}--\eref{alh-alh}) is an 
integrable model, which during its 40-year history have attracted considerable 
interest and which is one of the best-studied integrable systems. 

Thus, one can use various results that have been obtained for the HBDE 
and the ALH to derive 
solutions for \eref{eq-bist}--\eref{eq-alh} and hence for system 
\eref{eq-hcl-p} and \eref{eq-hcl-n}. Namely this approach is used in the 
following sections where we present two types of solutions: solution built of 
the Toeplitz determinants and the soliton solutions.

%%%%%%%%%%%%%%%%%%%%%%%%%%%%%%%%%%%%%%%%%%%%%%%%%%%%%%%%%%%%%%%%%%%%%%%%%%%%%%%% 
\subsection{Toeplitz solutions for the HL. \label{sec-tpl}} 
%%%%%%%%%%%%%%%%%%%%%%%%%%%%%%%%%%%%%%%%%%%%%%%%%%%%%%%%%%%%%%%%%%%%%%%%%%%%%%%% 

Here, we presents solutions for our field equations, constructed of 
the determinants of the Toeplitz matrices $\mytoeplitz{A}{m}{\ell}$, defined by 
\begin{equation}
  \mytoeplitz{A}{m}{\ell} = 
  \det\left| \omega_{m-a+b} \right|_{a,b=1,...,\ell} 
  \qquad 
  (\ell \ge 1)
\label{tpl-def-A}
\end{equation}
and $\mytoeplitz{A}{m}{0} = 1$.
It can be shown that these determinants satisfy the following identities:
\begin{equation}
  (\xi - \eta) 
  \mytoeplitz{A}{m+1}{\ell+1} 
  \myShiftedB{\xi\eta}{\mytoeplitz{A}{m}{\ell}} 
  = 
  \myShiftedB\xi{\mytoeplitz{A}{m}{\ell}} 
  \myShiftedB\eta{\mytoeplitz{A}{m+1}{\ell+1}} 
  - 
  \myShiftedB\xi{\mytoeplitz{A}{m+1}{\ell+1}} 
  \myShiftedB\eta{\mytoeplitz{A}{m}{\ell}} 
\label{tpl-Z}
\end{equation}
and 
\begin{equation}
  \myShiftedB\xi{\mytoeplitz{A}{m}{\ell}} 
  \myShiftedB\eta{\mytoeplitz{A}{m}{\ell}} 
  = 
  \mytoeplitz{A}{m}{\ell} 
  \myShiftedB{\xi\eta}{\mytoeplitz{A}{m}{\ell}} 
  + 
  \mytoeplitz{A}{m+1}{\ell+1} 
  \myShiftedB{\xi\eta}{\mytoeplitz{A}{m-1}{\ell-1}} 
\label{tpl-Y}
\end{equation}
with shifts $\myShift\zeta$ being defined by 
$
  \myShifted\zeta{\mytoeplitz{A}{m}{\ell}} = 
  \det\left| \myShifted\zeta{\omega_{m-a+b}} \right|_{a,b=1,...,\ell} 
$
where  
\begin{equation}
  \myShifted{\zeta}{\omega_{m}} 
  = 
  \omega_{m+1} - \zeta \omega_{m}.
\label{tpl-def-shift}
\end{equation}
We present, in \ref{app-toeplitz}, a sketch of a proof of these 
identities which is based on the results from \cite{V13}.

It is easy to see from \eref{tpl-Z} and \eref{tpl-Y} that tau-functions 
defined by 
\begin{equation}
  \sigma = \mytoeplitz{A}{m-1}{\ell-1}, 
  \quad
  \tau = \mytoeplitz{A}{m}{\ell}, 
  \quad
  \rho = \mytoeplitz{A}{m+1}{\ell+1} 
  \qquad
  ( \ell,m = \mbox{constant} )
\end{equation}
solve equations similar to \eref{alh-bist}--\eref{alh-alh} with 
$\myconst{a}{\xi,\eta} = \xi - \eta $ and $\myconst{b}{\xi,\eta} = 1 $, 
\begin{eqnarray}
  (\xi - \eta) 
  \tau 
  \myShiftedB{\xi\eta}{\sigma} 
  & = & 
  \myShiftedB\xi{\sigma} 
  \myShiftedB\eta{\tau} 
  - 
  \myShiftedB\eta{\sigma} 
  \myShiftedB\xi{\tau}, 
\\
  (\xi - \eta) 
  \rho 
  \myShiftedB{\xi\eta}{\tau} 
  & = & 
  \myShiftedB\xi{\tau} 
  \myShiftedB\eta{\rho} 
  - 
  \myShiftedB\eta{\tau} 
  \myShiftedB\xi{\rho}, 
\\
  \myShiftedB\xi{\tau} 
  \myShiftedB\eta{\tau} 
  & = &  
  \tau 
  \myShiftedB{\xi\eta}{\tau} 
  + 
  \rho 
  \myShiftedB{\xi\eta}{\sigma}. 
\end{eqnarray}
Thus, one can obtain solutions for our equations by identifying the 
translations by vectors $\vec{\alpha}_{i}$ with the shifts 
$\myShift{\alpha_{i}}$ where $\{ \alpha_{i} \}_{i=1..3}$ is a set of 
parameters. 
To write the final formulae, it is convenient to use the 
$n_{i}$-representation of the lattice vectors and the `Fourier' 
representation of the functions $\omega_{m}$, 
\begin{equation}
  \omega_{m} 
  =  
  \int\nolimits_{\gamma} dh \, 
  \hat\omega(h) \; 
  h^{m} 
\end{equation}
with arbitrary contour $\gamma$ and function $\hat\omega(h)$. The definition 
\eref{tpl-def-shift} of the shift $\myShift\zeta$ can be rewritten as 
\begin{equation}
  \myShifted{\zeta}{\omega_{m}} 
  = 
  \int\nolimits_{\gamma} dh \, 
  \hat\omega(h)  
  \left( h - \zeta \right)
  h^{m} 
\end{equation}
and can be extended, using \eref{def-nu-pm},  to 
\begin{equation}
  \omega_{m}(\vec{n}) 
  = 
  \int\nolimits_{\gamma} dh \, 
  \hat\omega(h, \vec{n}) \; 
  h^{m} 
\end{equation}
with 
\begin{equation}
  \hat\omega( h, \vec{n} ) 
  = 
  \hat\omega(h) \; 
  \prod_{i=1}^{3} 
  \left( h - \alpha_{i} \right)^{ 
    n_{i} - \mathcal{N}\left(n_{1},n_{2},n_{3}\right) 
  }
\qquad
  \biggl( \vec{n} = \sum_{i=1}^{3} n_{i}\vec{e}_{i}  \biggr)
\end{equation}
and $\mathcal{N}\left(n_{1},n_{2},n_{3}\right)$ being defined in 
\eref{def-N-cal}. 

Gathering the above formulae and making simple modifications, 
like setting $m=0$ 
(note that the factor $h^{m}$ can be incorporated into the definition of 
$\hat\omega(h)$) 
and introducing 
determinants $\mytoeplitz{D}{}{k}$ instead of 
$\mytoeplitz{A}{m+k}{\ell+k}$, 
we can formulate the following 
%
%%%%%%%%%%%%%%%%%%%%%%%%%%%%%%%%%%%%%%%%%%%%%%%%%%%%%%%%%%%%%%%%%%%%%%%%%%%%%%%%
\begin{proposition} \label{prop-toeplitz}
The Toeplitz solutions for the nonlinear HL equations 
\eref{eq-hcl-p}--\eref{eq-hcl-n} are given by  
\begin{equation}
  \myvar{u}(\vec{n}) = 
  \epsilon 
  \left[ 
    \frac{ \mytoeplitz{D}{}{-\epsilon}(\vec{n}) }
         { \mytoeplitz{D}{}{0}(\vec{n}) } 
  \right]^{\epsilon}
  \qquad 
  ( \vec{n} \in \Lambda^{\epsilon} )
\end{equation}
where $\epsilon = \pm 1$, $\Lambda^{\pm 1}=\Lambda^{\pm}$,  
\begin{equation}
  \mytoeplitz{D}{}{k}(\vec{n}) = 
  \left\{
  \begin{array}{lcl}
  \det\left| \omega_{k-a+b}(\vec{n}) \right|_{a,b=1,...,\ell+k} 
  &\quad&
  (\ell + k \ge 1) 
  \\
  1
  &&
  (\ell + k = 0) 
  \end{array}
  \right.
  \qquad 
  k = 0, \pm 1
\end{equation}
and 
\begin{equation}
  \omega_{m}(\vec{n}) 
  = 
  \int\nolimits_{\gamma} dh \, 
  \hat\omega(h) \; 
  h^{m} \, 
  \prod_{i=1}^{3} 
  \left( h - \alpha_{i} \right)^{ 
    n_{i} - \mathcal{N}\left(n_{1},n_{2},n_{3}\right) 
  }  
\qquad
  \biggl( \vec{n} = \sum_{i=1}^{3} n_{i}\vec{e}_{i}  \biggr)
\label{tpl-omega-n}
\end{equation}
with arbitrary positive integer $\ell$, parameters $\alpha_{i}$ ($i=1,2,3$), 
contour $\gamma$ and function $\hat\omega(h)$,  
and $\mathcal{N}\left(n_{1},n_{2},n_{3}\right)$ 
being defined in \eref{def-N-cal}. 

\end{proposition}
%%%%%%%%%%%%%%%%%%%%%%%%%%%%%%%%%%%%%%%%%%%%%%%%%%%%%%%%%%%%%%%%%%%%%%%%%%%%%%%%

Here, we would like to make a comment about the role of the construction 
\eref{eq-e-aa}. One can repeat the calculations of this section 
without introducing the vectors $\vec{\alpha}_{i}$ and using the shifts 
$ \myShift{\varepsilon_{i}} $ describing the `original' translations 
$ f(\vec{n}) \to f(\vec{n} + \vec{e}_{i})$. This leads to the relations 
\eref{tpl-omega-n} of proposition \ref{prop-toeplitz}  
with $\alpha_{i}$ replaced with $\varepsilon_{i}$ and $\mathcal{N}=0$. 
However, the restriction $ \prod_{i}\myShift{\varepsilon_{i}} = 1$ implies 
$ \prod_{i} \left( h - \varepsilon_{i} \right) = 1$ which means that the 
integral in \eref{tpl-omega-n} is reduced to the sum over the three 
roots of the cubic equation. Clearly, this family of solutions is 
noticeably less rich than the one presented above.

%%%%%%%%%%%%%%%%%%%%%%%%%%%%%%%%%%%%%%%%%%%%%%%%%%%%%%%%%%%%%%%%%%%%%%%%%%%%%%%%
\subsection{Soliton solutions for the HL. \label{sec-sls}}
%%%%%%%%%%%%%%%%%%%%%%%%%%%%%%%%%%%%%%%%%%%%%%%%%%%%%%%%%%%%%%%%%%%%%%%%%%%%%%%%

To derive the $N$-soliton solutions for our model we use the results of 
\cite{V15} where we have presented a large number of identities 
(soliton Fay identities) for the matrices of a special type. 
These $N{\times}N$ matrices are defined by 
\begin{equation}
  \begin{array}{lcl}
  \mymatrix{L} \mymatrix{A} - \mymatrix{A} \mymatrix{R} 
  & = & 
  \myunitket \mybra{a}, 
  \\
  \mymatrix{R} \mymatrix{B} - \mymatrix{B} \mymatrix{L} 
  & = & 
  \myunitket \mybra{b} 
  \end{array}
\label{eq-Sylvester} 
\end{equation}
where 
$\mymatrix{L}$ and $\mymatrix{R}$ are diagonal constant $N{\times}N$ matrices, 
$\myunitket$ is the $N$-column with all components equal to $1$, 
$\mybra{a}$ and $\mybra{b}$ are $N$-component rows that usually depend on the 
coordinates describing the model 
(note that we have replaced the $N$-columns 
$| \,\alpha\, \rangle$ and $| \,\beta\, \rangle$ used in \cite{V15} 
with  $\myunitket$, which can be done by means of the simple gauge transform).

In \cite{V15}, the soliton Fay identities are formulated in terms of the shifts 
defined by
\begin{equation}
  \begin{array}{lcl}
  \myShifted{\xi}{\mybra{a}} 
  & = & 
  \mybra{a} \left(\mymatrix{R} - \xi\right)^{-1}, 
  \\[2mm]
  \myShifted{\xi}{\mybra{b}} 
  & = & 
  \mybra{b} \left(\mymatrix{L} - \xi\right) 
  \end{array} 
\label{def-shits-ab}
\end{equation}
which determine the shifts of all other objects 
(the matrices $\mymatrix{A}$ and $\mymatrix{B}$, their determinants, 
the tau-functions constructed of $\mymatrix{A}$ and $\mymatrix{B}$ etc).

The soliton tau-functions have been defined in \cite{V15} as  
\begin{equation}
  \tau 
  = 
  \det \left| \mymatrix{1} + \mymatrix{A}\mymatrix{B} \right|
\end{equation}
and 
\begin{equation}
  \begin{array}{lcl}
  \sigma 
  & = & 
  \tau 
  \mybra{a} 
  ( \mymatrix{1} + \mymatrix{B}\mymatrix{A} )^{-1} 
  \myunitket, 
  \\[2mm] 
  \rho 
  & = & 
  \tau 
  \mybra{b} 
  ( \mymatrix{1} + \mymatrix{A}\mymatrix{B} )^{-1} 
  \myunitket
  \end{array} 
\label{sls-tau-sr} 
\end{equation}
The simplest soliton Fay identities, which are equations (3.12)--(3.14) 
of \cite{V15}, 
are exactly equations \eref{eq-alh} with $\myconst{b}{\xi,\eta} = 1 $ and 
\eref{eq-bist}--\eref{eq-bitr} with  $\myconst{a}{\xi,\eta} = \xi - \eta $. 

Thus, to obtain the $N$-soliton solutions one only needs to introduce the 
$\vec{n}$-dependence of the matrices $\mymatrix{A}$ and $\mymatrix{B}$ as well 
as of the rows $\mybra{a}$ and $\mybra{b}$, which can be done by 
\eref{def-nu-pm} and the identification of the 
translations by vectors $\vec{\alpha}_{i}$ with the shifts 
$\myShift{\alpha_{i}}$ ($i=1,2,3$): 
\begin{equation}
  \begin{array}{lcl} 
  \mymatrix{A}(\vec{n}) 
  & = & 
  \mymatrix{A}_{0} 
  \prod_{i=1}^{3} 
  \left( \mymatrix{R} - \alpha_{i} \right)^{ 
     \mathcal{N}\left(n_{1},n_{2},n_{3}\right) - n_{i} } 
  \\[2mm]
  \mymatrix{B}(\vec{n}) 
  & = & 
  \mymatrix{B}_{0} 
  \prod_{i=1}^{3} 
  \left( \mymatrix{L} - \alpha_{i} \right)^{ 
     n_{i} - \mathcal{N}\left(n_{1},n_{2},n_{3}\right) } 
  \end{array}
\label{sls-def-AB}
\end{equation}
with constant matrices $\mymatrix{A}_{0}$ and $\mymatrix{B}_{0}$ 
(that are, in fact, $\mymatrix{A}(\vec{0})$ and $\mymatrix{B}(\vec{0})$) 
and similar formulae for 
$\mybra{a(\vec{n})}$ and $\mybra{b(\vec{n})}$. 

After expressing, from \eref{eq-Sylvester},
$\mymatrix{A}_{0}$ and $\mymatrix{B}_{0}$ 
in terms of 
$\mybra{a_{0}} = \mybra{a(\vec{0})}$ 
and 
$\mybra{b_{0}} = \mybra{b(\vec{0})}$ 
one can write the resulting formulae as follows:
%
%%%%%%%%%%%%%%%%%%%%%%%%%%%%%%%%%%%%%%%%%%%%%%%%%%%%%%%%%%%%%%%%%%%%%%%%%%%%%%%%
\begin{proposition} \label{prop-solitons}
The $N$-soliton solutions for the nonlinear HL equations 
\eref{eq-hcl-p}--\eref{eq-hcl-n} are given by  
\begin{equation}
  \myvar{u}(\vec{n}) = 
  \mybra{c_{\pm} } 
  \left[ 
    \mymatrix{A}^{\mp 1}(\vec{n}) 
    + 
    \mymatrix{B}^{\pm 1}(\vec{n})  
  \right]^{-1}
  \myunitket^{\pm 1}
  \qquad 
  ( \vec{n} \in \Lambda^{\pm} ).
\end{equation}
where the matrices $\mymatrix{A}(\vec{n})$ and $\mymatrix{B}(\vec{n})$
are defined in \eref{sls-def-AB}
with 
\begin{equation}
  \mymatrix{A}_{0} 
  = 
  \left( 
    \frac{ a_{0k} }{ L_{j} - R_{k} }
  \right)_{j,k=1, ..., N}, 
\qquad
  \mymatrix{B}_{0} 
  = 
  \left( 
    \frac{ b_{0k} }{ R_{j} - L_{k} }
  \right)_{j,k=1, ..., N} 
\label{sls-def-AB0}
\end{equation}
and the constant N-rows $\mybra{c_{\pm}}$ are given by 
$ \mybra{c_{+}} = \mybra{1} \mymatrix{C} $
and 
$ \mybra{c_{-}} = \mybra{1} \mymatrix{C}^{T} $
(the superscript `T' stands for the  transposition) with 
\begin{equation}
  \mymatrix{C} = \mymatrix{M}^{-1}, 
  \qquad
  \mymatrix{M} 
  = 
  \left( 
    \frac{ 1 }{ L_{j} - R_{k} }
  \right)_{j,k=1, ..., N}. 
\end{equation}
Here, 
constants $\alpha_{i}$ ($i=1,2,3$), $L_{k}$, $R_{k}$, $a_{0k}$ and $b_{0k}$ 
($k=1, ...,N$) are the parameters describing the solution. 

\end{proposition}
%%%%%%%%%%%%%%%%%%%%%%%%%%%%%%%%%%%%%%%%%%%%%%%%%%%%%%%%%%%%%%%%%%%%%%%%%%%%%%%%

Again, as in section \ref{sec-tpl}, we would like to note that using the main 
result of this paper, formulated in proposition \ref{prop-bilin}, without 
introducing the $\vec{\alpha}$-vectors (see \eref{eq-e-aa}) one arrives at the 
restrictions
$ 
  \prod_{i=1}^{3} 
  \left( \mymatrix{L} - \varepsilon_{i} \right)
  = 
  \prod_{i=1}^{3} 
  \left( \mymatrix{R} - \varepsilon_{i} \right)
  = 
  \mymatrix{1}. 
$ 
Thus, for a given set of the $\varepsilon$-parameters, one has to 
construct two \emph{diagonal} matrices of only three roots of the cubic 
equation. 
Clearly, in this case one can hardly cross the Hirota's $3$-soliton threshold, 
whereas proposition \ref{prop-solitons}, resulting from \eref{eq-e-aa} gives 
$N$-soliton solutions for arbitrary $N$.

%%%%%%%%%%%%%%%%%%%%%%%%%%%%%%%%%%%%%%%%%%%%%%%%%%%%%%%%%%%%%%%%%%%%%%%%%%%%%%%%
\section{Exact solutions for the TL. \label{sec-exact-t}}
%%%%%%%%%%%%%%%%%%%%%%%%%%%%%%%%%%%%%%%%%%%%%%%%%%%%%%%%%%%%%%%%%%%%%%%%%%%%%%%%

In this section, we present solutions for equations \eref{eq-tri-p},
\begin{equation}
  \sum_{i=1}^{3}
    \Gamma_{i} \left[ 
    \frac{ 1 }{ \myvar{u}(\vec{n}) - \myvar{u}(\vec{n} + \vec{g}_{i}) } 
    +
    \frac{ 1 }{ \myvar{u}(\vec{n}) - \myvar{u}(\vec{n} - \vec{g}_{i}) }
    \right] 
  = 
  0
  \qquad  (\vec{n} \in \Lambda^{\mytriangle}) 
\label{eq-tri}
\end{equation}
that describe the nonlinear field model \eref{def-L-tri} defined on the TL 
\begin{equation}
  \Lambda^{\mytriangle} 
  = 
  \left\{
    \sum_{i=1}^{3} n_{i} \vec{g}_{i}, 
    \quad
    n_{i} \in \mathbb{Z} 
  \right\}. 
\end{equation}
In view of the results of section \ref{sec-def-t},  we do not need to perform 
any additional calculations. We can identify TL with $\Lambda^{+}$,
\begin{equation}
  \Lambda^{\mytriangle} 
  = 
  \Lambda^{+}
  = 
  \left\{ 
    \vec{n} = \sum\limits_{i=1}^{3} n_{i} \vec{e}_{i}
  \quad \Biggl| \quad 
  \sum\limits_{i=1}^{3} n_{i} = 0 \mathop{\mbox{mod}} 3 
  \right\},
\label{def-tri-latt}
\end{equation}
and use the formulae presented in the previous section. Here, we rewrite them 
in the vector form evading the use of the coordinates $n_{i}$. This can be 
done as follows. Any vector from $\Lambda^{+}$,  can be presented as
\begin{equation}
  \vec{n} = \sum\limits_{i=1}^{3} \hat{n}_{i} \vec{e}_{i}
\end{equation}
where the new integer coordinates, that satisfy 
$ \sum\nolimits_{i=1}^{3} \hat{n}_{i} = 0$, 
are given by  
\begin{equation}
  \hat{n}_{i} = n_{i} - \mathcal{N}\left(n_{1},n_{2},n_{3}\right) 
\end{equation}  
with integer (due to \eref{def-tri-latt})
$
  \mathcal{N}\left(n_{1},n_{2},n_{3}\right) 
  = 
  \frac{1}{3} 
  \sum\nolimits_{i=1}^{3} n_{i}. 
$
In the case of the equilateral lattice, 
\begin{equation}
  \vec{e}_{i}^{2} = e^{2}, 
  \qquad
  \angle\left(\vec{e}_{i},\vec{e}_{i+1} \right) = 2\pi/3 
  \qquad
  (i=1,2,3)
\end{equation}
the coordinates $\hat{n}_{i}$ can be expressed as 
\begin{equation}
  \hat{n}_{i} 
  = 
  \frac{2}{3 e^{2}} 
  \left( \vec{n}, \vec{e}_{i} \right) 
\end{equation}
where braces $( \, , )$ denote the standard scalar product. Thus, we can 
rewrite the coordinate-dependent formulae from the previous section in the 
vector form using the identity 
\begin{equation}
  \sum_{i=1}^{3} \hat{n}_{i} \lambda_{i} 
  = 
  \left( \vec{n}, \vec{\lambda} \right), 
  \qquad
  \vec{\lambda} 
  = 
  \frac{2}{3 e^{2}} 
  \sum_{i=1}^{3} \lambda_{i} \vec{e}_{i}. 
\end{equation}
In particular, sums that stem from the construction \eref{eq-e-aa} can be 
rewritten as follows: if 
\begin{equation}
  \alpha_{*} 
  = 
  \frac{1}{3} 
  \sum_{i=1}^{3} \alpha_{i}, 
\end{equation}
then 
\begin{equation}
  \sum_{i=1}^{3} n_{i} \left( \alpha_{i} - \alpha_{*}\right) 
  = 
  \frac{2}{3 e^{2}} 
  \left( 
    \vec{n}, 
    \sum_{i=1}^{3} \alpha_{i} \vec{e}_{i} 
  \right), 
\end{equation}
while the typical lattice factor, that appear in \eref{tpl-omega-n} 
and \eref{sls-def-AB}, can be written in the exponential form as  
\begin{equation}
  \prod_{i=1}^{3} 
  \left( x - \alpha_{i} \right)^{ 
    n_{i} - \mathcal{N}\left(n_{1},n_{2},n_{3}\right) 
  } 
  = 
  e^{ \left( \vec{\varphi}(x) , \vec{n}\right) } 
\end{equation}
where 
\begin{equation}
  \vec{\varphi}(x) 
  = 
  \frac{2}{3 e^{2}} 
  \sum_{i=1}^{3} 
  \vec{e}_{i} \, 
  \ln\left( x - \alpha_{i} \right). 
\label{def-phi}
\end{equation}
In what follows, we apply these vector formulae to modify the results of 
section \ref{sec-exact-h}.

%%%%%%%%%%%%%%%%%%%%%%%%%%%%%%%%%%%%%%%%%%%%%%%%%%%%%%%%%%%%%%%%%%%%%%%%%%%%%%%% 
\subsection{Toeplitz solutions for the TL. } 
%%%%%%%%%%%%%%%%%%%%%%%%%%%%%%%%%%%%%%%%%%%%%%%%%%%%%%%%%%%%%%%%%%%%%%%%%%%%%%%% 

It is clear that to obtain the Toeplitz solutions for equation \eref{eq-tri}, 
one can use results presented in proposition \ref{prop-toeplitz}. 
The only change that we make 
(except of adding the condition $\vec{n}\in\Lambda^{\mytriangle}=\Lambda^{+}$), 
is to the rewrite formula \eref{tpl-omega-n} for the $\vec{n}$-dependence of 
$\omega_{m}$. This leads to 
%
%%%%%%%%%%%%%%%%%%%%%%%%%%%%%%%%%%%%%%%%%%%%%%%%%%%%%%%%%%%%%%%%%%%%%%%%%%%%%%%%
\begin{proposition} \label{prop-toeplitz-t}
The Toeplitz solutions for the nonlinear TL equations 
\eref{eq-tri} are given by  
\begin{equation}
  \myvar{u}(\vec{n}) = 
  \frac{ \mytoeplitz{D}{}{-1}(\vec{n}) }
       { \mytoeplitz{D}{}{0}(\vec{n}) } 
  \qquad 
  ( \vec{n} \in \Lambda^{\mytriangle} )
\end{equation}
where 
\begin{equation}
  \mytoeplitz{D}{}{k}(\vec{n}) = 
  \left\{
  \begin{array}{lcl}
  \det\left| \omega_{k-a+b}(\vec{n}) \right|_{a,b=1,...,\ell+k} 
  &\quad&
  (\ell + k \ge 1) 
  \\
  1
  &&
  (\ell + k = 0) 
  \end{array}
  \right.
  \qquad 
  k = 0, -1
\end{equation}
and 
\begin{equation}
  \omega_{m}(\vec{n}) 
  = 
  \int\nolimits_{\gamma} dh \, 
  \hat\omega(h) \; 
  h^{m} \; 
  e^{ \left( \vec{\varphi}(h) , \vec{n}\right) } 
\end{equation}
with arbitrary positive constant $\ell$, parameters $\alpha_{i}$ ($i=1,2,3$), 
contour $\gamma$ and function $\hat\omega(h)$, and the vector function 
$\vec{\varphi}(h)$ being defined in \eref{def-phi}.

\end{proposition}
%%%%%%%%%%%%%%%%%%%%%%%%%%%%%%%%%%%%%%%%%%%%%%%%%%%%%%%%%%%%%%%%%%%%%%%%%%%%%%%%

%%%%%%%%%%%%%%%%%%%%%%%%%%%%%%%%%%%%%%%%%%%%%%%%%%%%%%%%%%%%%%%%%%%%%%%%%%%%%%%%
\subsection{Soliton solutions for the TL. }
%%%%%%%%%%%%%%%%%%%%%%%%%%%%%%%%%%%%%%%%%%%%%%%%%%%%%%%%%%%%%%%%%%%%%%%%%%%%%%%%

As in the case of the Toeplitz solutions, we use results obtained for the HL, 
gathered in proposition \ref{prop-solitons}, and present them as 
%
%%%%%%%%%%%%%%%%%%%%%%%%%%%%%%%%%%%%%%%%%%%%%%%%%%%%%%%%%%%%%%%%%%%%%%%%%%%%%%%%
\begin{proposition} 
\nopagebreak
The $N$-soliton solutions for the nonlinear TL equations \eref{eq-tri} 
are given by  
\begin{equation}
  \myvar{u}(\vec{n}) = 
  \mybra{1} \mymatrix{M}^{-1} 
  \left[ 
    \mymatrix{A}^{-1}(\vec{n}) 
    + 
    \mymatrix{B}(\vec{n})  
  \right]^{-1}
  \myunitket
  \qquad 
  ( \vec{n} \in \Lambda^{\mytriangle} ).
\end{equation}
where the matrices $\mymatrix{A}(\vec{n})$ and $\mymatrix{B}(\vec{n})$
are given by 
\begin{equation}
  \begin{array}{lclcl} 
  \mymatrix{A}(\vec{n}) 
  & = & 
  \mymatrix{A}_{0} \; 
  \mathrm{diag}\left( \; 
    e^{ - \left( \vec{\phi}_{k} , \vec{n}\right) } 
  \; \right)_{k=1, ..., N}
  &\qquad& 
  \vec{\phi}_{k} = \vec{\varphi}\left( R_{k} \right) %+ \vec{\phi}_{0k}
  \\[2mm]
  \mymatrix{B}(\vec{n}) 
  & = & 
  \mymatrix{B}_{0} \; 
  \mathrm{diag}\left( \; 
    e^{ \left( \vec{\psi}_{k} , \vec{n}\right) } 
  \; \right)_{k=1, ..., N} 
  &&
  \vec{\psi}_{k} = \vec{\varphi}\left( L_{k} \right) %+ \vec{\psi}_{0k}
  \end{array}
\end{equation}
with the vector function $\vec{\varphi}(h)$ defined in \eref{def-phi}, 
\begin{equation}
  \mymatrix{A}_{0} 
  = 
  \left( 
    \frac{ a_{0k} }{ L_{j} - R_{k} }
  \right)_{j,k=1, ..., N}, 
\qquad
  \mymatrix{B}_{0} 
  = 
  \left( 
    \frac{ b_{0k} }{ R_{j} - L_{k} }
  \right)_{j,k=1, ..., N} 
\end{equation}
and 
\begin{equation}
  \mymatrix{M} 
  = 
  \left( 
    \frac{ 1 }{ L_{j} - R_{k} }
  \right)_{j,k=1, ..., N}. 
\end{equation}

Here, 
constants $\alpha_{i}$ ($i=1,2,3$), $L_{k}$, $R_{k}$, 
%$\vec{\phi}_{0k}$ and $\vec{\psi}_{0k}$
$a_{0k}$ and $b_{0k}$ 
($k=1, ...,N$) are the parameters describing the solution. 
\end{proposition}

The one-soliton solution ($N=1$) can be presented, after replacing the vectors 
$\vec{\phi}_{1}$ and $\vec{\psi}_{1}$ with 
$\vec{\phi}_{\pm} = 
\frac{1}{2} \left( \vec{\psi}_{1} \pm \vec{\phi}_{1} \right)$ 
and the redefinition of the constants, in the following form:
\begin{equation}
  \myvar{u}(\vec{n}) = 
  \frac{ \exp\left[ 
         \delta_{+} - \left( \vec{\phi}_{+} , \vec{n}\right) 
         \right]} 
       { \cosh\left[ 
         \delta_{-} - \left( \vec{\phi}_{-} , \vec{n}\right) 
         \right]} 
\end{equation}
where 
\begin{equation}
  \vec{\phi}_{\pm} 
  = 
  \frac{1}{3 e^{2}} 
  \sum_{i=1}^{3} 
  \left[ 
    \ln\left( R - \alpha_{i} \right) \pm \ln\left( L - \alpha_{i} \right) 
  \right] 
  \vec{e}_{i}. 
\end{equation}
Here, $L$, $R$ (that we use instead of $L_{1}$ and $R_{1}$) and 
$\delta_{\pm}$ are arbitrary constants and 
$\alpha_{i}$ ($i=1,2,3$) are related to $\Gamma_{i}$ by 
$\Gamma_{i} = \alpha_{i+1} - \alpha_{i-1}$ 
(because of the arbitrariness of $L$ and $R$, 
one can take any particular solution of these equations, for example, 
$\alpha_{1} = - \Gamma_{2}$, $\alpha_{2} = \Gamma_{1}$ and $\alpha_{3} = 0$). 
Note that the presented solution is, at the same time, the 
$\Lambda^{+}$-part of the one-soliton solution for the HL 
(the values of $\myvar{u}(\vec{n})$ with $\vec{n}\in\Lambda^{-}$ can be 
obtained from \eref{eq-tri-a}).

%%%%%%%%%%%%%%%%%%%%%%%%%%%%%%%%%%%%%%%%%%%%%%%%%%%%%%%%%%%%%%%%%%%%%%%%%%%%%%%%
\section{Conclusion. }
%%%%%%%%%%%%%%%%%%%%%%%%%%%%%%%%%%%%%%%%%%%%%%%%%%%%%%%%%%%%%%%%%%%%%%%%%%%%%%%%

To conclude, we would like to give some comments related to the derivation of 
the explicit solutions presented in this paper, as well as to some questions 
that have not been discussed in the preceding sections. 

First we want to repeat the main steps that we made to take into account 
the specific features of the HL. The key moment of the bilinearization is the 
introduction of the tau-functions. One can see that in the equations of 
proposition \ref{prop-bilin} the tau-functions appear in a nonstandard way. 
The asymmetry between $\sigma$ and $\rho$ (the latter appears in the 
denominator) can be `explained' by the wish to obtain the homogeneity of the 
differences 
$\myvar{u}(\vec{n}) - \myvar{u}(\vec{n} \pm \vec{e}_{i})$ 
with respect to the scaling/gauge transformations 
$(\sigma,\tau,\rho) \to \left(e^{\lambda}\sigma,\tau,e^{-\lambda}\rho \right)$
which is typical in complex models, like the one by Ablowitz and Ladik, where, 
usually, one has to use the triplet of tau-functions.
As to the shift $\vec{\nu} \to \vec{\nu} - 2 \vec{\alpha}_{*}$ that we 
introduced in \eref{def-u-qr} for the negative sublattice $\Lambda^{-}$, 
we cannot give any sound `motivation', one can consider it as a trick to 
eliminate the asymmetry between the equations on $\Lambda^{+}$ and 
$\Lambda^{-}$. 

Now, we would like to return to the restriction \eref{restr-ge} or 
\eref{restr-gamma}. Restrictions of this type often appear in the studies of 
integrable models. If we consider, for example, the HBDE, 
the restriction similar to \eref{restr-gamma} is present in the most  
of the works devoted to this system (including the original paper \cite{H81}). 
However, as it has been demonstrated in, for example, \cite{RGS92}, it is not 
needed for integrability 
(the widespread opinion now is that it is required for the existence of 
Hirota-form soliton solutions). The same result one can find in, for example, 
\cite{ABS12}: in the list of the integrable systems the HBDE is written in the 
form that contradicts the Hirota's assumption.
Thus, one can try to obtain solutions for the model 
considered in this paper without \eref{restr-ge} or \eref{restr-gamma}
which is an interesting problem from the viewpoint of applications. 
The $\ln$-type potential (like in \eref{def-hcl-Le}) 
appears in physics as the two-dimensional Coulomb 
potential. However, \eref{restr-gamma} makes it impossible to interpret 
our model as describing a system of two-dimensional point charges 
because in this case the constants of interaction should be multiplicative, 
i.e. to be of the form 
$\Gamma_{i} \propto Q_{\vec{n}} Q_{\vec{n} \pm \vec{e}_{i}}$ 
($Q_{\vec{n}}$ is the value of the charge of the particle occupying a node at 
$\vec{n}$) with 
$\sum_{i=1}^{3} Q_{\vec{n} \pm \vec{e}_{i}} = 0$, 
and it is rather difficult to find examples of physical systems with such 
distribution of charges. 

Finally, we would like to point out possible continuations of this work. 
The most straightforward one is to consider the three-dimensional  
graphite-type lattices using the integrability of the enlarged set of 
equations \eref{eq-bist}--\eref{eq-alh}. 
Another possible generalization is to study time-dependent systems related to 
the action \eref{def-hcl-L} and \eref{def-hcl-Le}. 
However, these questions are out of the scope of the present paper and 
will be addressed in future publications.

\section*{Acknowledgments} 

We would like to thank the Referees for their constructive comments 
and suggestions for improvement of this paper.

\appendix

%%%%%%%%%%%%%%%%%%%%%%%%%%%%%%%%%%%%%%%%%%%%%%%%%%%%%%%%%%%%%%%%%%%%%%%%%%%%%%%% 
\section{Properties of the Hirota system. \label{app-hbde}} 
%%%%%%%%%%%%%%%%%%%%%%%%%%%%%%%%%%%%%%%%%%%%%%%%%%%%%%%%%%%%%%%%%%%%%%%%%%%%%%%% 

To simplify the presentation of the proofs of the statements made in section 
\ref{sec-alh} we introduce 
\begin{eqnarray}
    \mathfrak{s}_{\alpha,\beta} 
  & = & 
    \myconst{a}{\alpha,\beta}
    \myvar{\tau} 
    \myShiftedB{\alpha\beta}{\myvar{\sigma}} 
  - \myShiftedB{\alpha}{\myvar{\sigma}} 
    \myShiftedB{\beta}{\myvar{\tau}} 
  + \myShiftedB{\alpha}{\myvar{\tau}} 
    \myShiftedB{\beta}{\myvar{\sigma}}, 
\\
    \mathfrak{r}_{\alpha,\beta} 
  & = & 
    \myconst{a}{\alpha,\beta}
    \myvar{\rho} 
    \myShiftedB{\alpha\beta}{\myvar{\tau}} 
  - \myShiftedB{\alpha}{\myvar{\tau}} 
    \myShiftedB{\beta}{\myvar{\rho}} 
  + \myShiftedB{\alpha}{\myvar{\rho}} 
    \myShiftedB{\beta}{\myvar{\tau}}, 
\\
    \mathfrak{t}_{\alpha,\beta} 
  & = & 
    \myconst{b}{\alpha,\beta}
    \myShiftedB{\alpha}{\myvar{\tau}} 
    \myShiftedB{\beta}{\myvar{\tau}} 
  - \myvar{\tau} 
    \myShiftedB{\alpha\beta}{\myvar{\tau}} 
  - \myvar{\rho} 
    \myShiftedB{\alpha\beta}{\myvar{\sigma}} 
\label{app-eq-alh}
\end{eqnarray}
which are nothing but the right-hand sides of equations 
\eref{alh-bist}--\eref{alh-alh} (all gothic letters stand for the functions 
which are zero in the framework of the problem we study).

%%%%%%%%%%%%%%%%%%%%%%%%%%%%%%%%%%%%%%%%%%%%%%%%%%%%%%%%%%%%%%%%%%%%%%%%%%%%%%%% 
\subsection{Proof of \eref{alh-hbde} and \eref{alh-hbde-sr}. \label{app-hbde-1}} 

By trivial algebra one can show that the quantity 
\begin{equation}
    \mathfrak{t}'_{\alpha,\beta,\gamma} 
  = 
    \myconst{a}{\alpha,\beta}
    \myShiftedB{\gamma}{\myvar{\tau}} 
    \myShiftedB{\alpha\beta}{\myvar{\tau}} 
  - \myconst{a}{\alpha,\gamma}
    \myShiftedB{\beta}{\myvar{\tau}} 
    \myShiftedB{\alpha\gamma}{\myvar{\tau}} 
  + \myconst{a}{\beta,\gamma}
    \myShiftedB{\alpha}{\myvar{\tau}} 
    \myShiftedB{\beta\gamma}{\myvar{\tau}} 
\end{equation}
can be presented as a linear combination of $\mathfrak{r}_{\alpha,\beta}$:
\begin{equation}
    \myvar{\rho} \, 
    \mathfrak{t}'_{\alpha,\beta,\gamma} 
  = 
    \myShiftedB{\gamma}{\myvar{\tau}} 
    \mathfrak{r}_{\alpha,\beta} 
  - \myShiftedB{\beta}{\myvar{\tau}} 
    \mathfrak{r}_{\alpha,\gamma} 
  + \myShiftedB{\alpha}{\myvar{\tau}} 
    \mathfrak{r}_{\beta,\gamma} 
\end{equation}
This means that equations $\mathfrak{r}_{\alpha,\beta}=0$, i.e. equations 
\eref{alh-bitr} imply $\mathfrak{t}'_{\alpha,\beta,\gamma}=0$, 
i.e. equations \eref{alh-hbde}.
In a similar way, the functions 
\begin{eqnarray}
    \mathfrak{s}'_{\alpha,\beta,\gamma} 
  & = & 
    \myconst{a}{\alpha,\beta}
    \myShiftedB{\gamma}{\myvar{\sigma}} 
    \myShiftedB{\alpha\beta}{\myvar{\sigma}} 
  - \myconst{a}{\alpha,\gamma}
    \myShiftedB{\beta}{\myvar{\sigma}} 
    \myShiftedB{\alpha\gamma}{\myvar{\sigma}} 
  + \myconst{a}{\beta,\gamma}
    \myShiftedB{\alpha}{\myvar{\sigma}} 
    \myShiftedB{\beta\gamma}{\myvar{\sigma}} 
\\
    \mathfrak{r}'_{\alpha,\beta,\gamma} 
  & = & 
    \myconst{a}{\alpha,\beta}
    \myShiftedB{\gamma}{\myvar{\rho}} 
    \myShiftedB{\alpha\beta}{\myvar{\rho}} 
  - \myconst{a}{\alpha,\gamma}
    \myShiftedB{\beta}{\myvar{\rho}} 
    \myShiftedB{\alpha\gamma}{\myvar{\rho}} 
  + \myconst{a}{\beta,\gamma}
    \myShiftedB{\alpha}{\myvar{\rho}} 
    \myShiftedB{\beta\gamma}{\myvar{\rho}} 
\end{eqnarray}
can be presented as 
\begin{eqnarray}
    \myvar{\tau} \, 
    \mathfrak{s}'_{\alpha,\beta,\gamma} 
  & = & 
    \myShiftedB{\gamma}{\myvar{\sigma}} 
    \mathfrak{s}_{\alpha,\beta} 
  - \myShiftedB{\beta}{\myvar{\sigma}} 
    \mathfrak{s}_{\alpha,\gamma} 
  + \myShiftedB{\alpha}{\myvar{\sigma}} 
    \mathfrak{s}_{\beta,\gamma} 
\\
    \myShiftedB{\alpha\beta\gamma}{\myvar{\tau}} 
    \mathfrak{r}'_{\alpha,\beta,\gamma} 
  & = & 
    \myShiftedB{\alpha\beta}{\myvar{\rho}} 
    \myShiftedB{\gamma}{\mathfrak{r}_{\alpha,\beta}} 
  - \myShiftedB{\alpha\gamma}{\myvar{\rho}} 
    \myShiftedB{\beta}{\mathfrak{r}_{\alpha,\gamma}} 
  + \myShiftedB{\beta\gamma}{\myvar{\rho}} 
    \myShiftedB{\alpha}{\mathfrak{r}_{\beta,\gamma}} 
\end{eqnarray} 
which proves \eref{alh-hbde-sr}.

%%%%%%%%%%%%%%%%%%%%%%%%%%%%%%%%%%%%%%%%%%%%%%%%%%%%%%%%%%%%%%%%%%%%%%%%%%%%%%%% 
\subsection{Proof of \eref{alh-bt-st} and \eref{alh-bt-tr}. \label{app-hbde-2}} 
 
To prove \eref{alh-bt-st} and \eref{alh-bt-tr} we present the right-hand sides 
of the latter, 
\begin{eqnarray}
    \mathfrak{s}''_{\alpha,\beta,\gamma} 
  & = & 
    \myconst{a}{\alpha,\beta}
    \myShiftedB{\gamma}{\myvar{\tau}} 
    \myShiftedB{\alpha\beta}{\myvar{\sigma}} 
  - \myconst{a}{\alpha,\gamma}
    \myShiftedB{\beta}{\myvar{\tau}} 
    \myShiftedB{\alpha\gamma}{\myvar{\sigma}} 
  + \myconst{a}{\beta,\gamma}
    \myShiftedB{\alpha}{\myvar{\tau}} 
    \myShiftedB{\beta\gamma}{\myvar{\sigma}} 
\\
    \mathfrak{r}''_{\alpha,\beta,\gamma} 
  & = & 
    \myconst{a}{\alpha,\beta}
    \myShiftedB{\gamma}{\myvar{\rho}} 
    \myShiftedB{\alpha\beta}{\myvar{\tau}} 
  - \myconst{a}{\alpha,\gamma}
    \myShiftedB{\beta}{\myvar{\rho}} 
    \myShiftedB{\alpha\gamma}{\myvar{\tau}} 
  + \myconst{a}{\beta,\gamma}
    \myShiftedB{\alpha}{\myvar{\rho}} 
    \myShiftedB{\beta\gamma}{\myvar{\tau}} 
\end{eqnarray}
as
% e1
\begin{eqnarray}
    \myvar{\tau} \, 
    \mathfrak{s}''_{\alpha,\beta,\gamma} 
  & = & 
    \myShiftedB{\gamma}{\myvar{\tau}} 
    \mathfrak{s}_{\alpha,\beta} 
  - \myShiftedB{\beta}{\myvar{\tau}} 
    \mathfrak{s}_{\alpha,\gamma} 
  + \myShiftedB{\alpha}{\myvar{\tau}} 
    \mathfrak{s}_{\beta,\gamma} 
\\
    \myvar{\rho} \, 
    \mathfrak{r}''_{\alpha,\beta,\gamma} 
  & = & 
    \myShiftedB{\gamma}{\myvar{\rho}} 
    \mathfrak{r}_{\alpha,\beta} 
  - \myShiftedB{\beta}{\myvar{\rho}} 
    \mathfrak{r}_{\alpha,\gamma} 
  + \myShiftedB{\alpha}{\myvar{\rho}} 
    \mathfrak{r}_{\beta,\gamma} 
\end{eqnarray}

%%%%%%%%%%%%%%%%%%%%%%%%%%%%%%%%%%%%%%%%%%%%%%%%%%%%%%%%%%%%%%%%%%%%%%%%%%%%%%%% 
\subsection{Proof of \eref{alh-super-s}. \label{app-hbde-3}} 

The identity \eref{alh-super-s} follows from the fact that 
\begin{eqnarray}
    \mathfrak{S}_{\alpha,\beta,\gamma} 
  & = & 
    \myconst{a}{\alpha,\beta}
    \myconst{a}{\alpha,\gamma}
    \myconst{a}{\beta,\gamma}
    \myvar{\tau} 
    \myShiftedB{\alpha\beta\gamma}{\myvar{\sigma}} 
\nonumber \\
  &&
  - \myconst{a}{\beta,\gamma}
    \myShiftedB{\alpha}{\myvar{\sigma}} 
    \myShiftedB{\beta\gamma}{\myvar{\tau}} 
  + \myconst{a}{\alpha,\gamma}
    \myShiftedB{\beta}{\myvar{\sigma}} 
    \myShiftedB{\alpha\gamma}{\myvar{\tau}} 
  - \myconst{a}{\alpha,\beta}
    \myShiftedB{\gamma}{\myvar{\sigma}} 
    \myShiftedB{\alpha\beta}{\myvar{\tau}} 
\end{eqnarray}
satisfies 
\begin{equation}
    \myShiftedB{\gamma}{\myvar{\sigma}} 
    \mathfrak{t}'_{\alpha,\beta,\gamma} 
  + \myShiftedB{\gamma}{\myvar{\tau}} 
    \mathfrak{S}_{\alpha,\beta,\gamma} 
  =
    \myconst{a}{\alpha,\gamma}
    \myconst{a}{\beta,\gamma}
    \myvar{\tau} 
    \myShiftedB{\gamma}{\mathfrak{s}_{\alpha,\beta}} 
  + \myconst{a}{\beta,\gamma}
    \myShiftedB{\beta\gamma}{\myvar{\tau}} 
    \mathfrak{s}_{\alpha,\gamma} 
  - \myconst{a}{\alpha,\gamma}
    \myShiftedB{\alpha\gamma}{\myvar{\tau}} 
    \mathfrak{s}_{\beta,\gamma} 
\end{equation}
It should be noted that permutations of the parameters 
$\alpha$, $\beta$ and $\gamma$ lead to the same result 
(the left-hand side of the above equation is a skew-symmetric function). 
Thus, the equations \eref{alh-bist} for the three pairs of the parameters, 
$(\alpha,\beta)$, $(\alpha,\gamma)$ and $(\beta,\gamma)$ lead to the same 
expression for the function $\myShiftedB{\alpha\beta\gamma}{\myvar{\sigma}}$ 
which means that they have the property of the three-dimensional consistency.

%%%%%%%%%%%%%%%%%%%%%%%%%%%%%%%%%%%%%%%%%%%%%%%%%%%%%%%%%%%%%%%%%%%%%%%%%%%%%%%% 
\subsection{Derivation of \eref{alh-cab}. \label{app-hbde-4}} 

The necessity of the condition \eref{alh-cab} for the compatibility of 
\eref{alh-bist}--\eref{alh-alh} comes from the following observation: 
\begin{eqnarray}
&&
    \myvar{\tau} 
    \mathfrak{t}'_{\alpha,\beta,\gamma} 
  + \myvar{\rho} 
    \mathfrak{s}''_{\alpha,\beta,\gamma} 
  + \myconst{a}{\alpha,\beta}
    \myShiftedB{\gamma}{\myvar{\tau}} 
    \mathfrak{t}_{\alpha,\beta} 
  - \myconst{a}{\alpha,\gamma}
    \myShiftedB{\beta}{\myvar{\tau}} 
    \mathfrak{t}_{\alpha,\gamma} 
  + \myconst{a}{\beta,\gamma}
    \myShiftedB{\alpha}{\myvar{\tau}} 
    \mathfrak{t}_{\beta,\gamma} 
\nonumber \\&& \qquad
  = 
    \myShiftedB{\alpha}{\myvar{\tau}} 
    \myShiftedB{\beta}{\myvar{\tau}} 
    \myShiftedB{\gamma}{\myvar{\tau}} 
    \mathfrak{o}_{\alpha,\beta,\gamma} 
\end{eqnarray}
where $\mathfrak{o}_{\alpha,\beta,\gamma}$ is the right-hand side of equation 
\eref{alh-cab}.

%%%%%%%%%%%%%%%%%%%%%%%%%%%%%%%%%%%%%%%%%%%%%%%%%%%%%%%%%%%%%%%%%%%%%%%%%%%%%%%% 
\section{Equations \eref{alh-bist}--\eref{alh-alh} and the ALH. \label{app-alh}} 

To derive the ALH equations \eref{eq-alh-1} and \eref{eq-alh-2} we need the 
`deformed' version of \eref{alh-bist} and \eref{alh-bitr}, 
\begin{eqnarray}
    \myconst{a}{\alpha,\beta}
    \myvar{\tau} 
    \myShiftedB{\alpha\beta\gamma}{\myvar{\sigma}} 
  & = & 
  - \myconst{b}{\alpha,\gamma}
    \myShiftedB{\alpha}{\myvar{\tau}} 
    \myShiftedB{\beta\gamma}{\myvar{\sigma}} 
  + \myconst{b}{\beta,\gamma}
    \myShiftedB{\beta}{\myvar{\tau}} 
    \myShiftedB{\alpha\gamma}{\myvar{\sigma}} 
\label{app-alh-ts}
\\
    \myconst{a}{\alpha,\beta}
    \myvar{\rho} 
    \myShiftedB{\alpha\beta\gamma}{\myvar{\tau}} 
  & = & 
    \myconst{b}{\beta,\gamma}
    \myShiftedB{\beta}{\myvar{\rho}} 
    \myShiftedB{\alpha\gamma}{\myvar{\tau}} 
  - \myconst{b}{\alpha,\gamma}
    \myShiftedB{\alpha}{\myvar{\rho}} 
    \myShiftedB{\beta\gamma}{\myvar{\tau}} 
\label{app-alh-rt}
\end{eqnarray}
which can be obtained along the lines of \ref{app-hbde}. 
The above equations, 
\begin{eqnarray}
    \hat{\mathfrak{s}}_{\alpha,\beta,\gamma} 
  & = & 
    \myconst{a}{\alpha,\beta}
    \myvar{\tau} 
    \myShiftedB{\alpha\beta\gamma}{\myvar{\sigma}} 
  + \myconst{b}{\alpha,\gamma}
    \myShiftedB{\alpha}{\myvar{\tau}} 
    \myShiftedB{\beta\gamma}{\myvar{\sigma}} 
  - \myconst{b}{\beta,\gamma}
    \myShiftedB{\beta}{\myvar{\tau}} 
    \myShiftedB{\alpha\gamma}{\myvar{\sigma}} 
\\
    \hat{\mathfrak{r}}_{\alpha,\beta,\gamma} 
  & = & 
    \myconst{a}{\alpha,\beta}
    \myvar{\rho} 
    \myShiftedB{\alpha\beta\gamma}{\myvar{\tau}} 
  - \myconst{b}{\beta,\gamma}
    \myShiftedB{\beta}{\myvar{\rho}} 
    \myShiftedB{\alpha\gamma}{\myvar{\tau}} 
  + \myconst{b}{\alpha,\gamma}
    \myShiftedB{\alpha}{\myvar{\rho}} 
    \myShiftedB{\beta\gamma}{\myvar{\tau}} 
\end{eqnarray}
are the linear cobinations of the already used `zeroes':
\begin{eqnarray}
    \myShiftedB{\gamma}{\myvar{\tau}} 
    \hat{\mathfrak{s}}_{\alpha,\beta,\gamma} 
  & = & 
    \myvar{\tau} 
    \myShiftedB{\gamma}{\mathfrak{s}_{\alpha,\beta}} 
  + \myShiftedB{\beta\gamma}{\myvar{\sigma}} 
    \mathfrak{t}_{\alpha,\gamma} 
  - \myShiftedB{\alpha\gamma}{\myvar{\sigma}} 
    \mathfrak{t}_{\beta,\gamma} 
\\
    \myShiftedB{\alpha\beta}{\myvar{\tau}} 
    \hat{\mathfrak{r}}_{\alpha,\beta,\gamma} 
  & = & 
    \myShiftedB{\alpha\beta\gamma}{\myvar{\tau}} 
    \mathfrak{r}_{\alpha,\beta} 
  - \myShiftedB{\beta}{\myvar{\rho}} 
    \myShiftedB{\alpha}{\mathfrak{t}_{\beta,\gamma}} 
  + \myShiftedB{\alpha}{\myvar{\rho}} 
    \myShiftedB{\beta}{\mathfrak{t}_{\alpha,\gamma}} 
\end{eqnarray}
which means that equations 
$\mathfrak{s}_{\alpha,\beta} = 
 \mathfrak{r}_{\alpha,\beta} = 
 \mathfrak{t}_{\alpha,\beta}=0$ imply 
$\hat{\mathfrak{s}}_{\alpha,\beta,\gamma} = 
 \hat{\mathfrak{r}}_{\alpha,\beta,\gamma} = 0$.

But setting $\beta=\gamma=\kappa$ in \eref{app-alh-ts} and \eref{app-alh-rt}, 
applying $\myShift{\kappa}^{-1}$ and rewriting the resulting equations 
in terms of $Q$ and $R$ (see \eref{alh-def-QR} and \eref{alh-def-E}) 
one arrives at 
\begin{eqnarray}
    \myShift{\alpha} \myShift{\kappa}^{-1} \myvar{Q} 
  - \myvar{Q} 
  & = & 
    \myconst{a}{\alpha,\kappa}
    \myconst{b}{\alpha,\kappa} \; 
    \myvar{P_{\alpha}} \; 
    \myShift{\alpha} \myvar{Q} 
\\
    \myvar{R} 
  - \myShift{\alpha} \myShift{\kappa}^{-1} \myvar{R} 
  & = & 
    \myconst{a}{\alpha,\kappa}
    \myconst{b}{\alpha,\kappa} \; 
    \myvar{P_{\alpha}} \; 
    \myShift{\kappa}^{-1} \myvar{R} 
\end{eqnarray}
where 
\begin{equation}
  \myvar{P_{\alpha}} 
  =
  \frac{
    \left( \myShift{\alpha} \myvar{\tau} \right) 
    \left( \myShift{\kappa}^{-1} \myvar{\tau} \right) 
   }{
    \myconst{b}{\alpha,\kappa} \; 
    \myvar{\tau} 
    \left( \myShift{\alpha} \myShift{\kappa}^{-1} \myvar{\tau} \right) 
    }
\end{equation}
which can be presented (by means of the shifted by $\myShift{\kappa}^{-1}$ 
equations \eref{alh-alh} with $\beta=\kappa$ and 
the definitions of $Q$ and $R$) as 
\begin{equation}
  \myvar{P_{\alpha}} 
  = 
  1 
  - 
  \myvar{R} \left( \myShift{\alpha} \myShift{\kappa}^{-1} \myvar{Q} \right). 
\end{equation}
This completes the proof of \eref{eq-alh-1} and \eref{eq-alh-2}.

%%%%%%%%%%%%%%%%%%%%%%%%%%%%%%%%%%%%%%%%%%%%%%%%%%%%%%%%%%%%%%%%%%%%%%%%%%%%%%%% 
\section{Proof of \eref{tpl-Z} and \eref{tpl-Y}. \label{app-toeplitz}} 
%%%%%%%%%%%%%%%%%%%%%%%%%%%%%%%%%%%%%%%%%%%%%%%%%%%%%%%%%%%%%%%%%%%%%%%%%%%%%%%% 

Here, we present the outline of the derivation (or verification) of equations 
\eref{tpl-Z} and \eref{tpl-Y} for the Toeplitz determinants. 

We use some of the identities that where collected in \cite{V13} and that can 
be derived by applying the Jacobi identity to the determinants \eref{tpl-def-A} 
and the framed ones, 
\begin{equation}
  \mytoeplitz{F}{m}{\ell+1}(\zeta) = 
  \det\left|
  \begin{array}{ccccc} 
    1 & \zeta & \zeta^{2} & \dots & \zeta^{\ell} \cr
    \omega_{m-1} & \omega_{m} & \omega_{m+1} & \dots & \omega_{m + \ell -1} \cr
    \vdots & \vdots & \vdots & \ddots & \vdots
  \end{array} 
  \right|. 
\label{toeplitz-F}
\end{equation}
We start with equations (A.3) and (A.7) from the Appendix A of \cite{V13}: 
\begin{equation}
    \mytoeplitz{A}{k}{\ell-1}    \, \mytoeplitz{F}{k+1}{\ell+1}(\zeta) 
  - \mytoeplitz{A}{k+1}{\ell}    \, \mytoeplitz{F}{k}{\ell}(\zeta) 
  + \zeta\mytoeplitz{A}{k}{\ell} \, \mytoeplitz{F}{k+1}{\ell}(\zeta) 
  = 0 
\label{seZ3}
\end{equation}
and 
\begin{equation}
  \mytoeplitz{A}{k}{\ell} \mytoeplitz{F}{k}{\ell+1} 
  - \mytoeplitz{A}{k}{\ell+1} \mytoeplitz{F}{k}{\ell} 
  - \mytoeplitz{A}{k-1}{\ell} \mytoeplitz{F}{k+1}{\ell+1} 
  = 0. 
\label{seZ0}
\end{equation}
Rewriting $\mytoeplitz{F}{m}{\ell+1}$ as a $\ell$-determinant, 
\begin{equation}
  \mytoeplitz{F}{m}{\ell+1}(\zeta) 
  = 
  \det\left| 
    \Omega_{m+a-b}(\zeta) 
  \right|_{a,b=1,...,\ell} 
\end{equation} 
with
\begin{equation}
  \Omega_{m}(\zeta) = \omega_{m} - \zeta \omega_{m-1}. 
\end{equation}
and using definition \eref{tpl-def-shift} 
one can present $\mytoeplitz{F}{m}{\ell+1}$-determinants as 
shifted $\mytoeplitz{A}{m}{\ell}$-determinants, 
\begin{equation}
  \mytoeplitz{F}{m}{\ell+1}(\zeta)
  = 
  \myShifted\zeta{ \mytoeplitz{A}{m-1}{\ell} }
\end{equation}
which converts \eref{seZ3} and \eref{seZ0} into 
\begin{eqnarray}
  \mathfrak{p}^{m}_{\ell}(\zeta) 
  & := & 
  \zeta
  \mytoeplitz{A}{m}{\ell+1} 
  \myShiftedB\zeta{ \mytoeplitz{A}{m}{\ell} }
  - 
  \mytoeplitz{A}{m+1}{\ell+1} 
  \myShiftedB\zeta{ \mytoeplitz{A}{m-1}{\ell} }
  + 
  \mytoeplitz{A}{m}{\ell} 
  \myShiftedB\zeta{ \mytoeplitz{A}{m}{\ell+1} } 
  = 0,
\\
  \mathfrak{q}^{m}_{\ell}(\zeta) 
  & := & 
  \mytoeplitz{A}{m}{\ell} 
  \myShiftedB\zeta{ \mytoeplitz{A}{m+1}{\ell} }
  - 
  \mytoeplitz{A}{m+1}{\ell} 
  \myShiftedB\zeta{ \mytoeplitz{A}{m}{\ell} }
  + 
  \mytoeplitz{A}{m+1}{\ell+1} 
  \myShiftedB\zeta{ \mytoeplitz{A}{m}{\ell-1} } 
  = 0.
\end{eqnarray}
Now, we demonstrate that the combinations of the determinants that appear in 
\eref{tpl-Z} and \eref{tpl-Y}, 
\begin{equation}
  \mathfrak{u}^{m}_{\ell}(\xi,\eta) 
  := 
  (\xi-\eta)
  \mytoeplitz{A}{m+1}{\ell+1} 
  \myShiftedB{\xi\eta}{\mytoeplitz{A}{m}{\ell}} 
  - 
  \myShiftedB\xi{\mytoeplitz{A}{m}{\ell}} 
  \myShiftedB\eta{\mytoeplitz{A}{m+1}{\ell+1}} 
  + 
  \myShiftedB\xi{\mytoeplitz{A}{m+1}{\ell+1}} 
  \myShiftedB\eta{\mytoeplitz{A}{m}{\ell}} 
\end{equation}
and 
\begin{equation}
  \mathfrak{v}^{m}_{\ell}(\xi,\eta) 
  := 
  \myShiftedB\xi{\mytoeplitz{A}{m}{\ell}} 
  \myShiftedB\eta{\mytoeplitz{A}{m}{\ell}} 
  - 
  \mytoeplitz{A}{m}{\ell} 
  \myShiftedB{\xi\eta}{\mytoeplitz{A}{m}{\ell}} 
  - 
  \mytoeplitz{A}{m+1}{\ell+1} 
  \myShiftedB{\xi\eta}{\mytoeplitz{A}{m-1}{\ell-1}} 
\end{equation}
vanish by virtue of the equations
$\mathfrak{p}^{m}_{\ell}(\zeta)=\mathfrak{q}^{m}_{\ell}(\zeta)=0$.

It is a straightforward (though rather tedious) exercise in algebra to 
verify that 
\begin{equation}
  \begin{array}{l}
  \mytoeplitz{A}{m+1}{\ell} 
  \myShiftedB{\xi\eta}{\mytoeplitz{A}{m}{\ell-1}} 
  \mathfrak{u}^{m}_{\ell}(\xi,\eta) 
  \\[1mm]
  - 
  \mytoeplitz{A}{m+1}{\ell+1} 
  \myShiftedB{\xi\eta}{\mytoeplitz{A}{m}{\ell}} 
  \mathfrak{u}^{m}_{\ell-1}(\xi,\eta) 
  \end{array} 
  = 
  \begin{array}{l}
  \myShiftedB{\eta}{\mytoeplitz{A}{m}{\ell}} 
  \myShiftedB{\xi\eta}{\mytoeplitz{A}{m}{\ell-1}} 
  \mathfrak{p}^{m+1}_{\ell}(\xi)
  \\[1mm]
  - 
  \myShiftedB{\xi}{\mytoeplitz{A}{m}{\ell}} 
  \myShiftedB{\xi\eta}{\mytoeplitz{A}{m}{\ell-1}} 
  \mathfrak{p}^{m+1}_{\ell}(\eta)
  \\[1mm]
  - 
  \mytoeplitz{A}{m+1}{\ell+1} 
  \myShiftedB{\xi}{\mytoeplitz{A}{m+1}{\ell}} 
  \myShiftedB\eta{\mathfrak{p}^{m}_{\ell-1}(\xi)}
  \\[1mm]
  + 
  \mytoeplitz{A}{m+1}{\ell+1} 
  \myShiftedB{\eta}{\mytoeplitz{A}{m+1}{\ell}} 
  \myShiftedB\xi{\mathfrak{p}^{m}_{\ell-1}(\eta)} 
  \end{array}
\end{equation}
which means that 
\begin{equation}
  \mathfrak{u}^{m}_{\ell}(\xi,\eta) 
  = 
  u_{m}(\xi,\eta)
  \mytoeplitz{A}{m+1}{\ell+1} 
  \myShiftedB{\xi\eta}{\mytoeplitz{A}{m}{\ell}} 
\end{equation}
where the coefficients $u_{m}(\xi,\eta)$ do not depend on the sizes of the 
involved determinants. Thus, to determine $u_{m}(\xi,\eta)$ one can use the 
lowest-order version of this equation (say, with $\ell=1$) and obtain 
$u_{m}(\xi,\eta)=0$ and, hence, $\mathfrak{u}^{m}_{\ell}(\xi,\eta)=0$ 
which is the statement \eref{tpl-Z} that we want to prove.

In a similar way, it is straightforward to ascertain that 
\begin{equation}
\begin{array}{l}
  \myShiftedB{\xi}{\mytoeplitz{A}{m}{\ell}} 
  \myShiftedB{\eta}{\mytoeplitz{A}{m}{\ell}} 
  \mathfrak{v}^{m}_{\ell+1}(\xi,\eta)
  \\[1mm]
  - 
  \myShiftedB{\xi}{\mytoeplitz{A}{m}{\ell+1}}   
  \myShiftedB{\eta}{\mytoeplitz{A}{m}{\ell+1}}   
  \mathfrak{v}^{m}_{\ell}(\xi,\eta)
\end{array}
  = 
\begin{array}{l}
  - 
  \mytoeplitz{A}{m}{\ell+1} 
  \myShiftedB{\eta}{\mytoeplitz{A}{m}{\ell}} 
  \myShiftedB\xi{\mathfrak{p}^{m}_{\ell}(\eta)} 
  \\[1mm]
  - 
  \myShiftedB{\eta}{\mytoeplitz{A}{m}{\ell}} 
  \myShiftedB{\xi\eta}{\mytoeplitz{A}{m-1}{\ell}} 
  \mathfrak{q}^{m}_{\ell+1}(\xi)
  \\[1mm]
  + 
  \myShiftedB{\xi}{\mytoeplitz{A}{m}{\ell+1}} 
  \myShiftedB{\xi\eta}{\mytoeplitz{A}{m}{\ell}} 
  \mathfrak{p}^{m}_{\ell}(\eta)
  \\[1mm]
  + 
  \mytoeplitz{A}{m+1}{\ell+1} 
  \myShiftedB{\xi}{\mytoeplitz{A}{m}{\ell+1}} 
  \myShiftedB\eta{\mathfrak{q}^{m-1}_{\ell}(\xi)}\\[1mm]
\end{array}
\end{equation}
that leads to 
\begin{equation}
  \mathfrak{v}^{m}_{\ell}(\xi,\eta) 
  = 
  v_{m}(\xi,\eta)
  \myShiftedB{\xi}{\mytoeplitz{A}{m}{\ell}} 
  \myShiftedB{\eta}{\mytoeplitz{A}{m}{\ell}} 
\end{equation}
where the coefficient $v_{m}(\xi,\eta)$ is the same for all $\ell$. 
Again, rewriting the definition of  
$\mathfrak{v}^{m}_{\ell}(\xi,\eta)$ for small $\ell$ one can show that 
$v_{m}(\xi,\eta)$ and, hence, $\mathfrak{v}^{m}_{\ell}(\xi,\eta)$ are equal to 
zero, which proves \eref{tpl-Y}.

%%%%%%%%%%%%%%%%%%%%%%%%%%%%%%%%%%%%%%%%%%%%%%%%%%%%%%%%%%%%%%%%%%%%%%%%%%%%%%%%
%%%%%%%%%%%%%%%%%%%%%%%%%%%%%%%%%%%%%%%%%%%%%%%%%%%%%%%%%%%%%%%%%%%%%%%%%%%%%%%%
\end{document}